\pdfoutput=1

\documentclass[11pt,twoside,a4paper,cmspaper,final,collab]{cms-tdr}
\def\svnVersion{452982P}\def\cmsCernNoTag{CERN-EP-2017-264}\def\cmsCernDate{\today}\def\cmsMessage{Published in Physics Letters B as \href{http://dx.doi.org/10.1016/j.physletb.2018.03.019}{\doi{10.1016/j.physletb.2018.03.019.}}}

\begin{document}\cmsNoteHeader{EXO-16-003}

\hyphenation{had-ron-i-za-tion}
\hyphenation{cal-or-i-me-ter}
\hyphenation{de-vices}
\RCS$Revision: 452924 $
\RCS$HeadURL: svn+ssh://svn.cern.ch/reps/tdr2/papers/EXO-16-003/trunk/EXO-16-003.tex $
\RCS$Id: EXO-16-003.tex 452924 2018-03-27 17:26:23Z hardenbr $
\newlength\cmsFigWidth
\ifthenelse{\boolean{cms@external}}{\setlength\cmsFigWidth{0.85\columnwidth}}{\setlength\cmsFigWidth{0.4\textwidth}}
\ifthenelse{\boolean{cms@external}}{\providecommand{\cmsLeft}{upper\xspace}}{\providecommand{\cmsLeft}{left\xspace}}
\ifthenelse{\boolean{cms@external}}{\providecommand{\cmsRight}{lower\xspace}}{\providecommand{\cmsRight}{right\xspace}}
\ifthenelse{\boolean{cms@external}}{\providecommand{\suppMaterial}{the supplemental material  [URL will be inserted by publisher]\xspace}}
{\providecommand{\suppMaterial}{Appendix~\ref{app:suppMat}\xspace}}
\ifthenelse{\boolean{cms@external}}{\providecommand{\suppMaterialii}{the supplemental material\xspace}}
{\providecommand{\suppMaterialii}{Appendix~\ref{app:suppMat}\xspace}}
\providecommand{\NA}{\ensuremath{\text{---}}}
\providecommand{\HT}{\ensuremath{H_\mathrm{T}}\xspace}

\cmsNoteHeader{EXO-16-003}
\title{Search for new long-lived particles at $\sqrt{s} = 13$\TeV}

\date{\today}

\abstract{ A search for long-lived particles was performed with data corresponding to
 an integrated luminosity of 2.6\fbinv collected at a center-of-mass energy of
13\TeV by the CMS experiment in 2015.  The analysis exploits two customized topological
 trigger algorithms, and uses the  multiplicity of displaced jets to search for
 the presence of a signal  decay occurring at distances
 between 1 and 1000\mm. The results can be interpreted in a variety of different models. For pair-produced long-lived particles decaying
 to two b quarks and two leptons with equal decay rates between lepton flavors,  cross
 sections larger than 2.5\unit{fb} are excluded for proper decay lengths between 70--100\mm for a
 long-lived particle mass of 1130\GeV at 95\% confidence.
 For a specific model of pair-produced, long-lived top squarks with
 R-parity violating decays to a b quark and a lepton, masses below
 550--1130\GeV are excluded at 95\% confidence for equal branching fractions between lepton flavors, depending on the squark decay length.
This mass bound is the most stringent to date for top squark proper decay lengths greater than 3\mm.}

\hypersetup{
pdfauthor={CMS Collaboration},
pdftitle={Search for new long-lived particles at sqrt(s) = 13 TeV},
pdfsubject={CMS},
pdfkeywords={CMS, physics, long-lived particles}}

\maketitle
\section{Introduction}

The observation of physics beyond the standard model (BSM) is one of the main
objectives of the ATLAS and CMS experiments at the CERN LHC. With no
signal yet observed, these experiments have placed stringent bounds on
BSM models. The majority of these searches focus on particles with lab frame
decay lengths of $c\tau <1$\unit{mm} and incorporate selection requirements that
 reject longer-lived particle
decays. This leaves open the possibility that long-lived particles
could be produced but remain undetected. The present analysis exploits
information originating from the CMS calorimeters to reconstruct jets
and measure their energies. The information from reconstructed tracks,
in particular the transverse impact parameter, is used to
discriminate the signal of a jet whose origin is
displaced with respect to the primary vertex, from the background of
ordinary multijet events. The analysis is performed on data from proton-proton
collisions at $\sqrt{s}=13$\TeV, collected with the
CMS detector in 2015. The data set corresponds to an
integrated luminosity of 2.6\fbinv. Results for similar
signatures at $\sqrt{s}=8$\TeV have been reported by
ATLAS~\cite{PhysRevD.92.012010,Aad2015,Aaboud:2017iio},
CMS~\cite{CMS:2014wda},  and LHCb~\cite{Aaij:2014nma,Aaij:2016isa}. In this Letter, we present a new, more
general approach to searching for long-lived particles decaying to
combinations of jets and leptons, which is inclusive in event topology and does not require the reconstruction
of a  displaced vertex.

\section{The CMS detector}
\label{sec:detector}

The central feature of the CMS apparatus is a superconducting solenoid
of 6\unit{m} internal diameter, providing a magnetic field of
3.8\unit{T}. Within the solenoid volume are a silicon pixel and strip
tracker, a lead tungstate crystal electromagnetic calorimeter (ECAL), and a
brass and scintillator hadron calorimeter (HCAL), each composed of a barrel
and two endcap sections. Forward calorimeters extend the
pseudorapidity ($\eta$) coverage provided by the barrel and endcap
detectors. Muons are measured in gas-ionization detectors embedded in
the steel flux-return yoke outside the solenoid.

The silicon tracker measures charged particles with
$\abs{\eta} < 2.5$. It consists of silicon
pixels and silicon strip detector modules. The innermost pixel (strip)
layer is at a radial distance of 4.3 (44) \unit{cm} from the beamline.

The ECAL consists of lead tungstate
crystals and provides coverage in $\abs{ \eta } <
1.48 $ in a barrel region (EB) and $1.48 < \abs{ \eta } < 3.0$ in
two endcap regions (EE). A preshower detector composed of two planes
of silicon sensors interleaved with a total of 3 radiation lengths of lead is
located in front of the EE. The inner face
of the ECAL is at a radial distance of 129 \unit{cm} from the beamline.

In the region $\abs{ \eta } < 1.74$, the HCAL cells have widths of
0.087 in pseudorapidity and 0.087 radians in azimuth ($\phi$). In the
$\eta$-$\phi$ plane, and for $\abs{\eta} < 1.48$, the HCAL cells map
onto $5 \times 5$ arrays of ECAL crystals to form calorimeter towers
projecting radially outwards from close to the nominal interaction
point. For $1.74 < \abs{ \eta } < 3.00$, the coverage of the towers increases
progressively to a maximum of 0.174 in $\Delta \eta$ and $\Delta
\phi$. Within each tower, the energy deposits in ECAL and HCAL cells
are summed to define the calorimeter tower energies and are subsequently used
to provide the energies of jets. The inner face of the HCAL is at a radial distance
of 179 \unit{cm} from the beamline.

For each event, jets are clustered from energy deposits in the calorimeters,
using the \textsc{FastJet}~\cite{fastjet} implementation of the
anti-\kt algorithm~\cite{Cacciari:2008gp}, with
the distance parameter 0.4. Tracks that are within
 $\Delta R = \sqrt{\smash[b]{(\Delta \eta)^2 + (\Delta \phi)^2}}< 0.4$ of
 a jet are considered to be associated with the jet.

Events of interest are selected using a two-tiered trigger system~\cite{Khachatryan:2016bia}. The first level, composed of custom hardware processors, uses information from the calorimeters and muon detectors to select events at a rate of around 100\unit{kHz} within a time interval of less than 4\mus. The second level, known as the high-level trigger (HLT), consists of a farm of processors running a version of the full event reconstruction software optimized for fast processing, and reduces the event rate to around 1\unit{kHz} before data storage.

A more detailed description of the CMS detector, together with a
definition of the coordinate system used and the relevant kinematic
variables, can be found in Ref.~\cite{cmsdet}.

\section{Data sets and simulated samples}
\label{sec:samples}
Events are selected using two dedicated HLT algorithms,
designed to identify events with displaced jets. Both algorithms have a
requirement on $\HT$, which is defined as the scalar sum of
the transverse momentum $\pt$ of the jets in the event, considering only jets with $\pt > 40\GeV$ and $\abs{\eta} < 3.0$.
The inclusive algorithm accepts events with $\HT >500\GeV$ and at least
two jets, each with $\pt>40\GeV$, $\abs{\eta}<2.0$, and no more than two associated prompt tracks.
Tracks are classified as prompt if their transverse impact
parameter relative to the beam line, $\mathrm{IP}^{\mathrm{2D}}$, is less than 1\unit{mm}. The exclusive algorithm requires $\HT>350\GeV$ and at least
two jets with $\pt>40\GeV$, $\abs{\eta} < 2.0$, no more than two associated
prompt tracks, and at least one associated track with $\mathrm{IP}^{\mathrm{2D}}>5\sigma_{\mathrm{IP}^{\mathrm{2D}}}$,
where $\sigma_{\mathrm{IP}^{\mathrm{2D}}}$ is the calculated uncertainty in
$\mathrm{IP}^{\mathrm{2D}}$.  Data collected by algorithms with
identical $\HT$ requirements and no tracking requirements
 are used to study the performance of the online selection algorithms.

Events are selected offline by requiring at least two jets with
$\pt>60\GeV$ and $\abs{\eta}<2.0$. Two classes of events are
considered: events (i) passing the inclusive online algorithm and
with $\HT>650\GeV$ and (ii) passing the exclusive
online algorithm and with $\HT>450\GeV$. Combining these
two classes of events results in 786\,002 unique events. We
refer to these events as passing the event selection or simply
``Selection'' in the efficiency tables.

The main source of background events originates from multijet
production. The properties of this background process are studied
using a simulated multijet sample, generated with
$\MADGRAPH 5$ \cite{Alwall:2011uj} and interfaced with $\PYTHIA8$
\cite{Sjostrand:2007gs} for parton showering and hadronization.
The NNPDF 2.3~\cite{NNPDF23}
parton distribution functions (PDFs) are used to model the parton
momentum distribution inside the colliding protons. The event
simulation includes the effect of additional proton-proton collisions in
the same bunch crossing and in bunch crossings nearby in time, referred
to as pileup. Simulated samples are reweighted to match the pileup
profile observed in data. The detector response is
simulated in detail using \GEANTfour~\cite{GEANT4}.

The analysis is interpreted with a set of benchmark signal models.
The {Jet-Jet} model predicts pair-produced long-lived
scalar neutral particles $X^{0}$, each decaying to a quark-antiquark pair, where
possible pairs include \cPqu,
\cPqd, \cPqs, \cPqc, and \cPqb~quarks.  The two scalars are produced through a
$2\to 2$ scattering process, mediated by a $\cPZ^*$
propagator, and the decay rate to each flavor is assumed to be the same. The resonance
mass $m_{X^0}$ and average proper decay length $c\tau_0$ are varied between
50 and 1500\GeV and between 1 and 2000\mm, respectively. The model resembles hidden
valley models that produce long-lived neutral final states \cite{ScalarX}. The
trigger efficiencies for $m_{X^0}=300\GeV$ and
$c\tau_0 = 1, 30,$ and 1000\mm are 30\%, 81\%, and 42\%,
 respectively. For example, the trigger efficiencies are 2\%, 14\%,
and 92\% for $c\tau_0 = 30\mm$ and $m_{X^{0}} = 50, 100,$
and 1000\GeV respectively. The trigger efficiency is calculated from the total
number of events passing only the logical OR of the two trigger paths.

The {B-Lepton} model contains pair-produced long-lived top squarks
in R-parity \cite{Farrar:1978xj} violating models of supersymmetry (SUSY)~\cite{Graham:2012th}.
Each top squark decays to one \cPqb~quark and a lepton, with equal
decay rates to each of the three lepton flavors. The
resonance mass $m_{\sTop}$ and proper decay length $c\tau_0$ are
varied between 300 and 1000\GeV and between 1 and 1000\mm,
respectively. For example, the trigger efficiencies for
$m_{\sTop}=300\GeV$ and $c\tau_0 = 1, 30,$ and 1000\mm
 are 15\%, 41\%, and 23\%, respectively. The trigger efficiencies are 64\%, 71\%, and 74\%
for  $c\tau_0=30$ \mm and  $m_{\sTop} = 500, 700,$ and 1000\GeV, respectively.

Variations of these models with modified branching fractions are also investigated.
The {Light-Light} model is the Jet-Jet
model excluding decays to b quarks (equal decays to lighter quarks)
and the {B-Muon}, {B-Electron}, and {B-Tau} models are
derived from the B-Lepton model with 100\% branching fraction to
muons, electrons, and $\tau$ leptons, respectively. Both leptonic and hadronic
$\tau$ lepton decays are included
in the {B-Tau} interpretation. All signal samples are
generated with $\PYTHIA8$, with the same configuration as for the
multijet sample.

\section{Event selection and inclusive displaced-jet tagger}
\label{sec:tagger}

In general, events contain multiple primary vertex (PV) candidates, corresponding to pileup collisions occurring in the same proton bunch crossing.  The PV reconstruction employs Gaussian constraints
 on the reconstructed position based on the luminous region, which
 is evaluated from the reconstructed PVs in many events.
A description of the PV reconstruction can be found in Ref.~\cite{pvreco}. The displaced-jet
 identification variables utilize the
PV with the highest $\pt^2$ sum of the constituent tracks.
The results of the analysis are found to be insensitive to the choice of
 the method used to select the PV, since the uncertainty in the
 transverse position of the primary vertex is small relative to the signal model decay lengths.

The analysis utilizes a dedicated tagging algorithm to identify
displaced jets. For each jet, the algorithm takes as input the
reconstructed tracks within $\Delta R < 0.4$ of the jet. All tracks with $\pt>1\GeV$ that are
 selected by all iterations of track reconstruction are considered. A detailed
list of requirements for the CMS track collection can be found elsewhere \cite{pvreco}.
 Three variables are considered for each jet in the event.
The first variable quantifies how likely it is that the jet
originates from a given PV\@. For a given jet,  $\alpha_\text{jet} (\mathrm{PV})$
is defined for each PV as
\begin{equation}
\alpha_{\text{jet}}(\mathrm{PV}) = \frac{\sum_{\text{tracks} \in \mathrm{PV}} \pt^{\text{tracks}}}{\sum_{\text{tracks}} \pt^{\text{tracks}}},
\end{equation}
where the sum in the denominator is over all tracks associated with the jet and the sum
in the numerator is over just the subset of these tracks originating from the given PV\@.
The tagging variable $\alpha_{\text{max}}$ is the largest value of
$\alpha_{\text{jet}}(\mathrm{PV})$ for the jet.

The second variable quantifies the significance of the measured
transverse displacement for the jet. For each track associated with the
jet, the significance of the track's transverse impact parameter,
$\mathrm{IP}^{\mathrm{2D}}_{\text{sig}}$, is computed as the ratio of the
track's $\mathrm{IP}^{\mathrm{2D}}$ and its uncertainty.  The tagging
variable $\widehat{\mathrm{IP}}{}^{\mathrm{2D}}_{\text{sig}}$ is the median of the
$\mathrm{IP}^{\mathrm{2D}}_{\text{sig}}$ distribution of all tracks in a jet.

The third variable quantifies the angular difference between the emission angle of a given
track in a jet and the parent particle flight direction.  For
each track associated with the jet, $\Theta_{\mathrm{2D}}$ is computed as the angle between
the track $\vec{p_{\text{T}}}=(p_{\text{x}},p_{\text{y}})$ at
the track's innermost hit and the vector connecting the chosen PV to
this hit in the transverse plane. The tagging variable
$\widehat{\Theta}_{\mathrm{2D}}$ is the median of the
$\Theta_{\mathrm{2D}}$ distribution for the tracks associated with the
jet.

It should be noted that leptons giving rise to calorimeter energy deposits (tau leptons and electrons) will also be classified as ``displaced jets'', if the associated track(s) satisfies the tagging criteria, and thus contribute to the search sensitivity. Additionally, by not requiring the reconstruction of a displaced vertex,
 the analysis is becomes sensitive to pair-produced long-lived decays
 with a single reconstructed track per decay.

\begin{figure*}
\centering
\includegraphics[width=.32\textwidth]{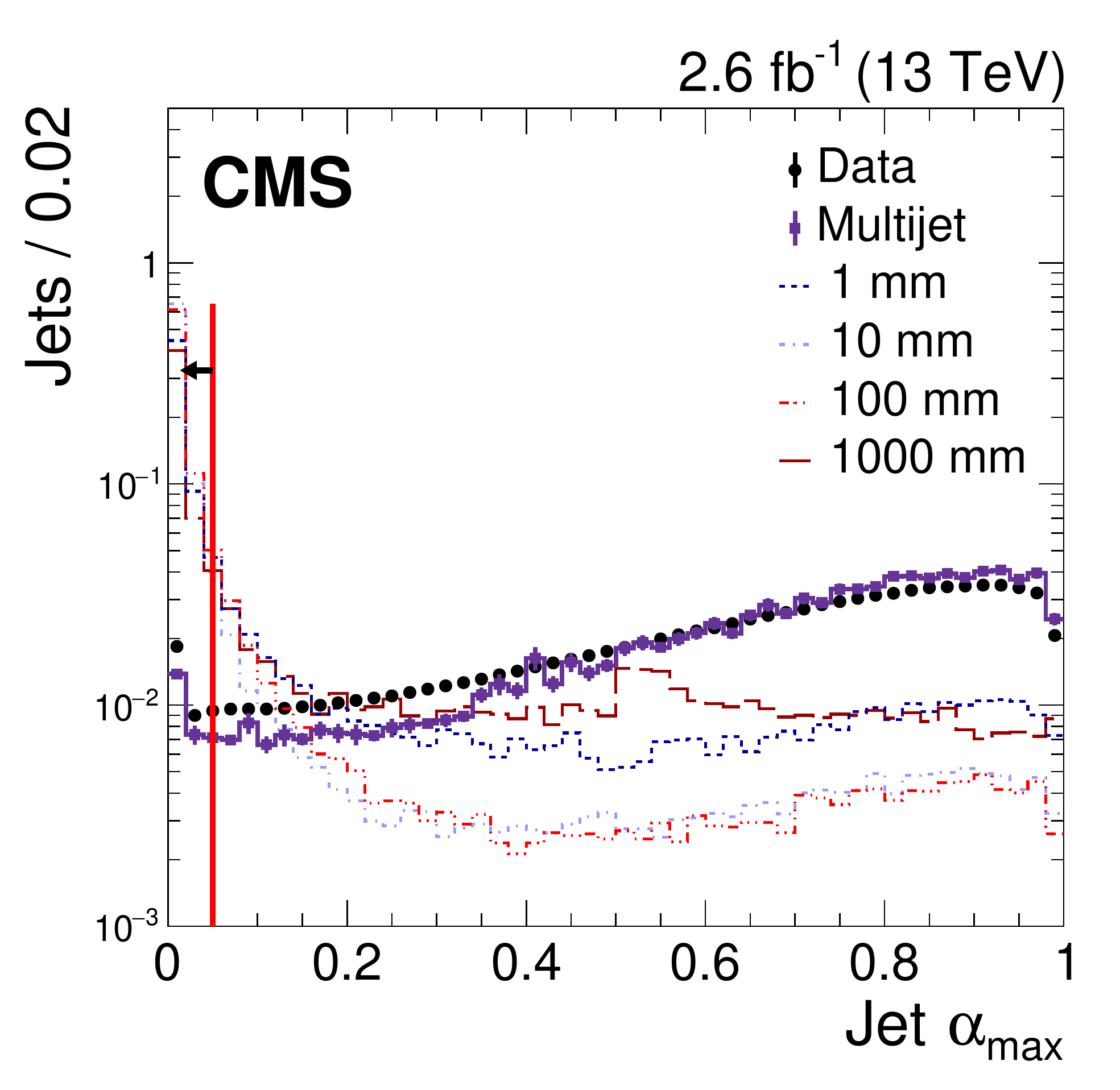}
\includegraphics[width=.32\textwidth]{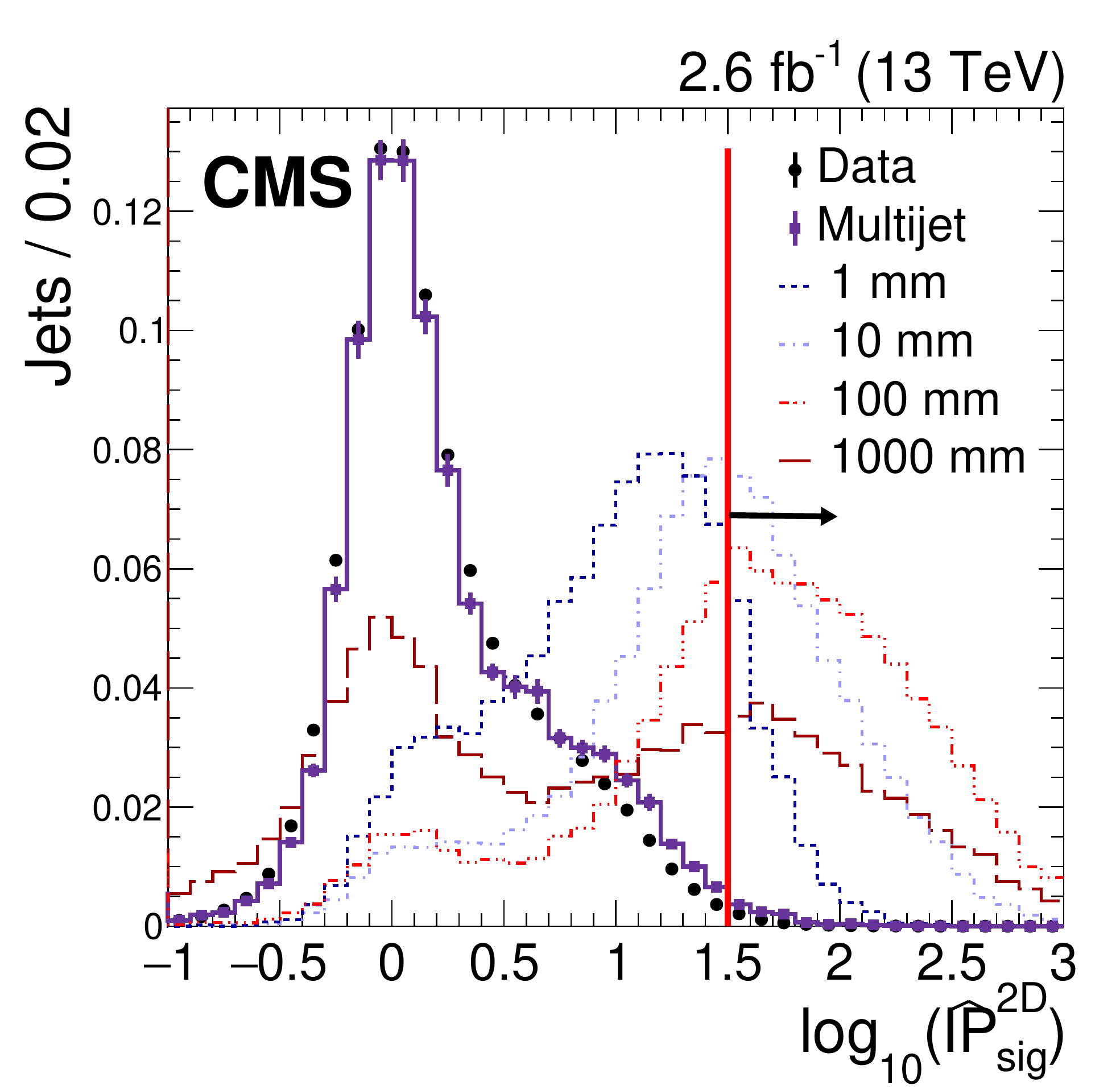}
\includegraphics[width=.32\textwidth]{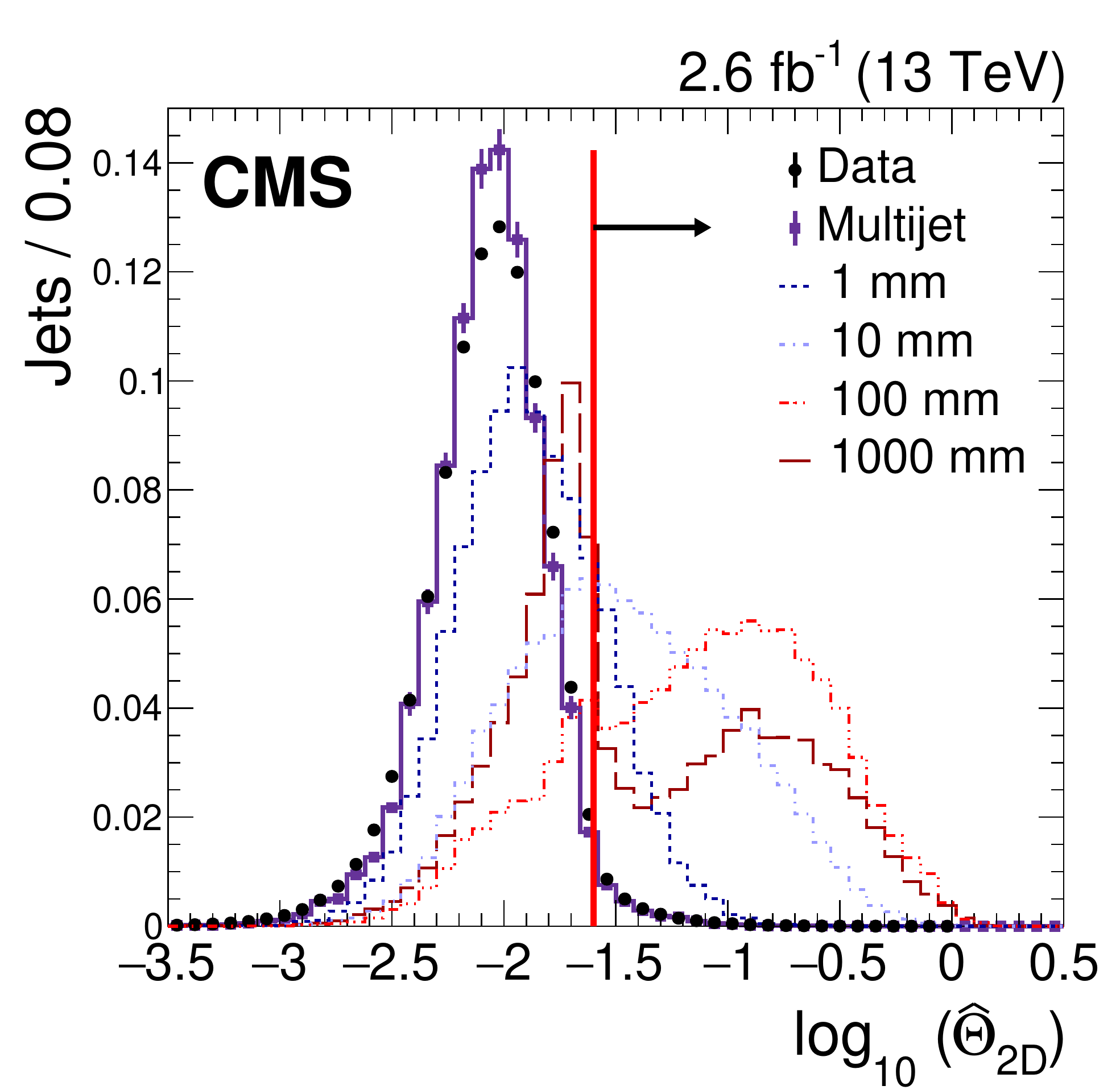}
\caption{Comparison of distributions for the displaced-jet
  tagging variables $\alpha_{\text{max}}$ (left),
  $\widehat{\mathrm{IP}}{}^{\mathrm{2D}}_{\text{sig}}$ (center), and
  $\widehat{\Theta}_{\mathrm{2D}}$ (right) in data and simulation.
  The data distributions
  (circles) are compared to the expected background distributions from
  multijet events (squares) and several Jet-Jet benchmark
  models (dotted histograms) of pair-produced long-lived neutral
  scalar particles with $m_{X^{0}} = 700 \GeV$ and different values of
  $c\tau_0$.  The vertical lines designate the value of
  the requirement for the chosen displaced-jet tag. The direction of the
  arrow indicates the values included in the
  requirement. All distributions have unit normalization. \label{fig:varDist}}.
\end{figure*}

Figure~\ref{fig:varDist} shows the distributions of the three tagging
variables for data events, simulated multijet events, and simulated
signal events with $m_{X^0}=700\GeV$ and several values of $c\tau_0$.
Note that any mismodeling resulting from the multijet background does not affect
the analysis because the background estimate is derived from data. Simulation
of the multijet background only describes misidentified displaced jets.

The displaced-jet identification criteria are
$\alpha_{\text{max}}<0.05$,
$\log_{10}(\widehat{\mathrm{IP}}{}^{\mathrm{2D}}_{\text{sig}})> 1.5$, and
$\log_{10}(\widehat{\Theta}_{\mathrm{2D}})> -1.6$.  This selection was
chosen by selecting parameters that yielded the best discovery sensitivity for
the Jet-Jet model across all generated decay lengths and masses.

The average displaced-jet tagging efficiency with no trigger selection applied
 for $m_{X^0}=700\GeV$ is
4\% for $c\tau_0=1\mm$, 57\% for $c\tau_0=30\mm$, and 33\%
for $c\tau_0=1000\mm$. For $c\tau_0 >1000\mm$, the
long-lived particles typically decay beyond the tracker.
 For $c\tau_0 < 3\mm$, the experimental signature
for signal events becomes increasingly difficult to distinguish from that
of background b quark jets.

The search is performed by applying the selection criteria described above
and by  counting the number of tagged displaced
jets, $N_{\text{tags}}$.  In addition to the online and offline
requirements described in Section~\ref{sec:samples}, the analysis
signal region requires $N_{\text{tags}} \geq 2$.
Efficiencies are reported for the Jet-Jet and B-Lepton models as a function of
decay length with fixed mass (Table \ref{tab:cutflow_300gev}) as well as a function of
mass with fixed decay length (Table \ref{tab:cutflow_30mm}). Efficiencies for the
Light-Light, B-Tau, B-Electron, and B-Mu models are included in \suppMaterial as Tables
\ifthenelse{\boolean{cms@external}}{1}{\ref{tab:cutflow_BR_mass}} and \ifthenelse{\boolean{cms@external}}{2}{\ref{tab:cutflow_BR_lifetime}}.

\begin{table*}[tb]
\centering
  \topcaption{ Signal efficiencies (in \%) for $m_{X^0}=m_{\sTop}=300\GeV$
    for various values of $c\tau_0$ for the Jet-Jet and B-Lepton models. Selection requirements  are cumulative from
    the first row to the last. \label{tab:cutflow_300gev}}
\begin{tabular}{lcccc}
 \multicolumn{5}{c}{{Jet-Jet}} \\
 \hline
 $c\tau_0$\,[mm] & 1 & 10 & 100 & 1000 \\
 \hline
 $\geq$2 tags & 2.33 $\pm$ 0.15 & 39.49 $\pm$ 0.63 & 54.54 $\pm$ 0.74 & 14.58 $\pm$ 0.38 \\
 Trigger & 2.16 $\pm$ 0.15 & 38.12 $\pm$ 0.62 & 39.32 $\pm$ 0.63 & 8.07 $\pm$ 0.28 \\
Selection& 2.09 $\pm$ 0.14 & 37.09 $\pm$ 0.61 & 36.53 $\pm$ 0.60 & 6.67 $\pm$ 0.26 \\
 $\geq$3 tags &       0.17 $\pm$ 0.04 &       14.14 $\pm$ 0.38 &   16.72 $\pm$ 0.41 &    1.36 $\pm$ 0.12 \\
 $\geq$4 tags &       0.01 $\pm$ 0.01 &       4.73 $\pm$ 0.22 &    4.71 $\pm$ 0.22 &     0.17 $\pm$ 0.04 \\
\end{tabular}

\vspace*{3ex}
\centering
\begin{tabular}{ccccc}
\multicolumn{5}{c}{{B-Lepton}} \\
 \hline
$c\tau_0 \,[\text{mm}]$ & 1 & 10 & 100 & 1000 \\
\hline
$\geq$2 tags & 0.45 $\pm$ 0.02 & 15.82 $\pm$ 0.13 & 31.52 $\pm$ 0.19 & 8.55 $\pm$ 0.10   \\
Trigger       & 0.29 $\pm$ 0.02 & 11.45 $\pm$ 0.11 & 17.08 $\pm$ 0.14 & 3.22 $\pm$ 0.06   \\
Selection     & 0.27 $\pm$ 0.02 & 9.91 $\pm$ 0.11  & 13.33 $\pm$ 0.12 & 2.08 $\pm$ 0.05   \\
$\geq$3 tags & 0.02 $\pm$ 0.01 & 2.46 $\pm$ 0.05  & 3.81 $\pm$ 0.07  & 0.37 $\pm$ 0.02   \\
$\geq$4 tags & \NA              & 0.30 $\pm$ 0.02  & 0.48 $\pm$ 0.02  & 0.03 $\pm$ 0.01 \\
\end{tabular}
\end{table*}
\begin{table*}[tb]
\centering
  \topcaption{ Signal efficiencies (in \%) for the Jet-Jet and B-Lepton models
    with $c\tau_0= 30\mm$ and for various values of mass. Selection requirements are cumulative from
    the first row to the last. \label{tab:cutflow_30mm}}

\begin{tabular}{lccccc}
\multicolumn{6}{c}{{Jet-Jet}} \\
 \hline
 $m_{X^0}\,[\GeVns]$ & 50 & 100 & 300 & 1000 & 1500 \\
 \hline
 $\geq$2 tags & 2.71 $\pm$ 0.10 & 14.80 $\pm$ 0.22  & 54.24 $\pm$ 0.74 & 79.93 $\pm$ 0.89 & 82.55 $\pm$ 0.91 \\
 Trigger       & 0.50 $\pm$ 0.04 & 5.39 $\pm$ 0.13   & 46.41 $\pm$ 0.68 & 74.05 $\pm$ 0.86 & 77.65 $\pm$ 0.88 \\
 Selection     & 0.30 $\pm$ 0.03 & 3.70 $\pm$ 0.11   & 44.75 $\pm$ 0.67 & 73.99 $\pm$ 0.86 & 77.53 $\pm$ 0.88 \\
 $\geq$3 tags & 0.05 $\pm$ 0.01 & 1.09 $\pm$ 0.10   & 20.87 $\pm$ 0.46 & 49.42 $\pm$ 0.70 & 55.28 $\pm$ 0.74 \\
 $\geq$4 tags & \NA             & 0.22 $\pm$ 0.03   & 6.81 $\pm$ 0.26  & 25.45 $\pm$ 0.50 & 32.26 $\pm$ 0.57 \\
\end{tabular}

\vspace*{3ex}
\begin{tabular}{lcccc}
\multicolumn{5}{c}{{B-Lepton}}\\
\hline
 $m_{\sTop}\,[\GeVns]$ & 300 & 600 & 800 & 1000 \\
 \hline
 $\geq$2 tags & 31.52 $\pm$ 0.19  & 47.32 $\pm$ 0.23  & 52.53 $\pm$ 0.24  & 55.88 $\pm$ 0.35  \\
Trigger        & 17.08 $\pm$ 0.14  & 35.03 $\pm$ 0.20  & 40.40 $\pm$ 0.21  & 43.14 $\pm$ 0.30  \\
Selection      & 14.70 $\pm$ 0.13  & 32.34 $\pm$ 0.19  & 36.94 $\pm$ 0.20  & 39.26 $\pm$ 0.29  \\
$\geq$3 tags  & 4.11 $\pm$ 0.07 & 10.76 $\pm$ 0.11  & 13.29 $\pm$ 0.12  & 15.00 $\pm$ 0.18  \\
$\geq$4 tags  & 0.55 $\pm$ 0.03 & 1.83 $\pm$ 0.05   & 2.69 $\pm$ 0.05 & 3.09 $\pm$ 0.08 \\
\end{tabular}
\end{table*}

\section{Background prediction}
\label{sec:bkg}

Background events arise from jets containing tracks that
are mismeasured as displaced and jets containing tracks from the weak
decays of strange, charm, and bottom hadrons.

\begin{figure}
\centering
\includegraphics[width=.495\textwidth]{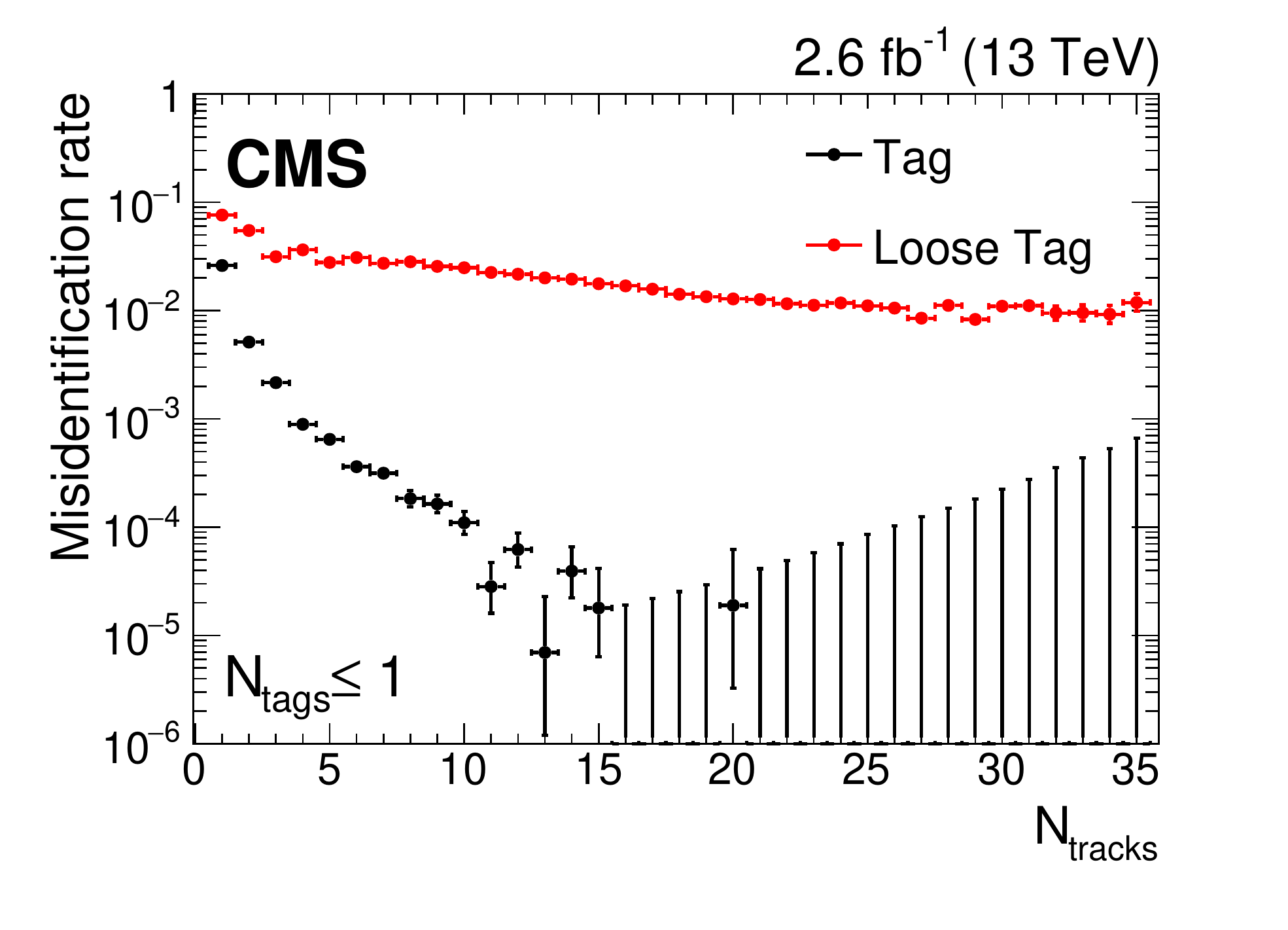}
\caption{The fraction of jets passing the displaced-jet tagging criteria as a function of the number of
tracks associated with the jet.
The results are from data events with $N_{\text{tags}} \leq 1$, collected with the displaced-jet triggers and passing the offline selection criteria. \label{fig:fake_rate}}
\end{figure}

To maintain the statistical independence of the events
that are used to perform the prediction and the events in the signal
region, the misidentification rate is
measured in a control sample defined as events with
 $N_{\text{tags}}\leq 1$ (as shown in Fig.~\ref{fig:fake_rate}), while the  signal region requires
$N_{\text{tags}} \geq 2$. Additionally, this control sample definition limits signal contamination.
There are 1391 events in data with $N_{\text{tags}}= 1$. The size of the bias introduced by only measuring the misidentification rate in
events with $N_\text{tags}\leq 1$ is quantifiable. For the chosen tag
requirement,
the effect of removing events with $N_\text{tags}>1$ on
 the predicted number of two tag events is negligible (0.4\%)
compared to the statistical uncertainty of the prediction.

Since the proportion of tracks identified as being displaced
is small and approximately constant, the likelihood of tagging a
 nondisplaced jet as a displaced jet decreases approximately
exponentially with the number of tracks
associated with the jet, $N_\text{tracks}$. Figure~\ref{fig:fake_rate} shows the fraction of jets that are tagged as displaced jets in data as a function of $N_\text{tracks}$.
 This function is the misidentification rate of tagging a prompt jet as displaced (assuming no signal contamination)  and is interpreted as the probability  $p(N_\text{tracks})$ of being tagged. This
 parameterization allows an event by event estimation of the probability of
 tagging any multiplicity of displaced jets.

Because of the high jet production cross section, even though the misidentification
rate is small, events with one tagged displaced jet are completely
dominated by standard model backgrounds, and signal contamination
can be ignored, even if the associated cross section is large.
This is explicitly verified with signal injection tests,
which are discussed below.

The misidentification rate is used to predict the probability $P(N_{\text{tags}})$ for an event to have $N_\text{tags}$ tagged jets. For instance, for an event $m$ with three jets $j_1$, $j_2$, and $j_3$, there is one jet configuration with no tags,
with a probability:
\begin{equation*}
P^m(N_{\text{tags}}=0) = (1-p_1)(1-p_2)(1-p_3),
\end{equation*}
where $p_i = p(N_{\text{tracks}}(j_i))$.  Similarly, there are three
jet configurations for this same event to have $N_{\text{tags}}=1$:
\ifthenelse{\boolean{cms@external}}{
\begin{multline*}
P^m(N_{\text{tags}}=1) = p_1(1-p_2)(1-p_3)
+  (1-p_1)p_2(1-p_3)\\
+  (1-p_1)(1-p_2)p_3.
\end{multline*}
}{
\begin{equation*}
P^m(N_{\text{tags}}=1) = p_1(1-p_2)(1-p_3)
+  (1-p_1)p_2(1-p_3)
+  (1-p_1)(1-p_2)p_3.
\end{equation*}
}
The probability of finding $N_{\text{tags}}$  tags in the $m$ event is:
\begin{equation}
P^m(N_{\text{tags}}) = \sum_{\text{jet-configs}} \prod_{i\in
  \text{tagged}} p_i \prod_{k \in \text{nontagged}} (1-p_k).
\label{eq:ntag_prob}
\end{equation}
Tagged jets enter the product as $p_i$ and nontagged jets enter as
$(1-p_i)$. Equation~(\ref{eq:ntag_prob}) is used to compute $P^m(N_{\text{tags}})$,
 under the assumption that the sample does not contain any signal. The number of events
expected for a given value of $N_{\text{tags}}$ is computed as
\begin{equation}
N_{\text{events}}(N_{\text{tags}}) = \sum_{m} P^m(N_{\text{tags}}),
\end{equation}
where $m$ runs only over events with fewer than two tagged jets.  The prediction is then compared
to the observed $N_{\text{tags}}$ multiplicity in events with two or
more tagged jets, to assess the presence of a signal.

We validate this procedure in the absence (background-only test) and
presence (signal injection test) of a signal, using simulated events.

The background-only test is performed by predicting the tag multiplicity
from the simulated multijet sample, using the distribution obtained for the misidentification rate.
In order to populate the large-$N_{\text{tags}}$
region of the distribution, a looser version of the displaced-jet
tagger is employed in this test. The loose displaced-jet identification criteria are
$\alpha_{\text{max}}<0.5$,
$\log_{10}(\widehat{\mathrm{IP}}{}^{\mathrm{2D}}_{\text{sig}})> 0.4$, and
$\log_{10}(\widehat{\Theta}_{\mathrm{2D}})> -1.7$.  The average misidentification rate of the loose (chosen)
tag definition is 2.6\%\,(0.05\%). The loose definition requirements were relaxed until
 a minimal number of two tag events were available to perform the background-only test.
The full sample of events passing the
event selection is divided into multiple independent samples and the background
prediction validated. The predicted background of $N_{\text{tags}}$ events
in simulated multijet events is found to be consistent with the observed number of events. The
associated pull distributions are found to have mean 0 and variance 1
as expected in the ideal case.

The signal injection test is performed by adding events of
pair-produced resonances decaying to two jets to the multijet sample
and repeating the procedure described above. In this case, the
chosen displaced-jet tagger is used.  The injected signal has  $m_{X^0} = 700\GeV$
and $c\tau_0 = 10\mm$ with a cross section varied in the range from 30\unit{fb} to 3\unit{pb}.  The jet probability
is computed as in the data, where no prior knowledge of the nature
 of the events (signal or background) is available. In this case,
 the misidentification rate is derived from the
mixed sample itself, including the contamination from the injected
signal sample. The signal contamination is found to have a minimal impact
on the predicted number of events in the signal region. For example, with an injected signal
cross section of  30\unit{fb}, 19  events are observed with two tags, while the two tag
prediction is consistent with the predictions obtained
for zero injected events: $N_{\text{events}}(N_{\text{tags}}\geq 2)=1.3$.
As another example, with an injection signal cross section of 3 pb,
 no three tag events are predicted,
while 1520 events with three tags are observed.
Given the insensitivity of the predicted background to
large amounts of injected signal, the analysis is robust to signal contamination
of the control region.

\section{Systematic uncertainties}
\label{sec:sys}

\subsection{Background systematic uncertainties}
\label{sec:bkgsys}

There is an uncertainty in the estimated background level associated with the choice of method used.
 This uncertainty is evaluated by repeating the background prediction procedure
described in Section 5 using the looser version of the displaced-jet
 tagging algorithm. The result is compared with that obtained using the nominal
 method and the observed difference of 7.5\% is taken as the systematic
 uncertainty from this source. This value for the uncertainty is used
 also for the three or more tags case.

The statistical uncertainty in the measured misidentification rate as a
function of $N_{\text{tracks}}$ is propagated to the predicted
$N_{\text{tags}}$ distribution as a systematic uncertainty. This systematic uncertainty
is calculated for each tag multiplicity bin. The uncertainty for the two tag
bin is 13\%.

\subsection{Signal systematic uncertainties}
\label{sec:sigsys}

All signal systematic uncertainties are calculated individually for
each model, for each mass and decay length point, and for each value
 of $N_\text{tags}$ in the signal region. In cases where the
 uncertainty depends on the mass, decay length, and/or decay mode
 of the long-lived particle, a range is quoted, referring to
 the uncertainty for $N_\text{tags}=2$ events.  A summary of the
 systematic uncertainties associated with the signal is given in Table \ref{tab:sigSys}.

The uncertainty in the trigger emulation is measured by comparing
the predicted efficiency
for simulated multijet events with that measured for data collected with a loose
$\HT$ trigger. The observed difference at the offline
$\HT$ threshold (5\%) is taken as an estimate of the uncertainty in the emulation of the
online $\HT$ requirement.  Similarly, the uncertainty
induced by the online versus offline jet acceptance is obtained from
the shift in the trigger efficiency when the offline minimum jet \pt
requirement is increased from 60 to 80\GeV\,(5\%).

The systematic uncertainty in the modeling of the online tracking
efficiency is obtained by studying the online regional track
reconstruction in data and in simulation. The online values of $\mathrm{IP}^{\mathrm{2D}}$
and $\mathrm{IP}^{\mathrm{2D}}_{\text{sig}}$ are varied by the magnitude of
the mismodeling found in events collected by control sample triggers consisting of
only an \HT requirement ($\HT> 425$ and \HT$> 275$).  The
new values are used to determine if the event would still pass at
least one of the trigger paths and its associated offline
\HT requirement. The $N_{\text{tags}}$ distribution is
recalculated with the values varied up and down. The relative change
in the number of events per $N_\text{tags}$ bin is taken as the systematic
uncertainty. For $N_{\text{tags}}=2$, this uncertainty varies from
1 to 35\%.

The systematic uncertainty in the luminosity is 2.3\% \cite{LUMI}.

The uncertainty arising from the choice of PDFs for pair-produced particles with
masses in the range of 50--1500\GeV is found to be 1--6\%.  An ensemble of
alternative PDFs is sampled from the output of the NNPDF fit.  Events
are reweighted according to the ratio between these alternative PDF
sets and the nominal ones. The distribution of the signal prediction
for these PDF ensembles is used to quantify this uncertainty.

The systematic uncertainty in the modeling of the jet tagging
variables in the signal simulation samples is estimated from the displaced track
modeling in multijet events in data and simulation. The mismodeling of the
measured value of $\Theta_{\mathrm{2D}}$ and $\mathrm{IP}_{\text{sig}}^{\mathrm{2D}}$ for
single tracks is propagated to the final tag distribution by varying
the individual measured values in simulation by the difference in the measured
value relative to data (3--10\%). The tagging variables are then
recalculated.  The $N_{\text{tags}}$ distribution is recalculated
with the new values. The systematic uncertainty is assigned as the
relative change in the number of events for each $N_{\text{tags}}$ bin. For the
two tag bin, this varies from 1 to 30\% depending on the mass and
decay length. The mismodeling of $\alpha_{\text{max}}$ is found to have
a negligible effect on the signal efficiency, as the requirement is
relatively loose.

\begin{table}[tb]
  \topcaption{Summary of the signal systematic uncertainties.
    When the uncertainty depends on the specific features of the models
    (mass, decay length, and decay mode of the long-lived particle) a
    range is quoted, which refers to the computed uncertainty for $N_{\text{tags}}=2$ events.~\label{tab:sigSys}}
\centering
\begin{tabular}{lc}
{Signal systematic uncertainty} & {Effect on yield} \\
\hline
$\HT$ trigger inefficiency &  5\% \\
Jet \pt trigger inefficiency  & 5\% \\
Trigger online tracking modeling & 1--35\% \\
Integrated Luminosity & 2.3\% \\
Acceptance due to the PDF choice & 1--6\% \\
Displaced-jet tag variable modeling & 1--30\% \\
\end{tabular}
\end{table}

\section{Results and interpretation}
\label{sec:results}

\begin{table}[tb]
  \topcaption{The predicted and observed number of events as a function
    of the number of tagged displaced jets. The prediction is based on the
    misidentification rate derived from events with fewer than two tags.
    The full event selection is applied. The uncertainty
    corresponds to the total background systematic uncertainty.\label{tab:result}}
\centering
\begin{tabular}{lcc}
{$N_{\text{tags}}$} & {Expected} & {Observed} \\
\hline
2 & $1.09\pm 0.16$ & 1  \\
$\geq$3 & $(4.9 \pm 1.0) \times 10^{-4}$ & 0 \\
\end{tabular}
\end{table}

The numerical values for the expected and observed yields are
summarized in Table~\ref{tab:result}.  The observed yields are found to
be consistent with the predicted background, within the statistical
and systematic uncertainties. No evidence for a signal at large values
of $N_{\text{tags}}$ is observed.

Exclusions for each model are obtained from the predicted and observed
event yields in Table~\ref{tab:result} and the signal efficiencies in
Tables~\ref{tab:cutflow_300gev} and \ref{tab:cutflow_30mm} and \ifthenelse{\boolean{cms@external}}{Tables  1 and 2}{Tables \ref{tab:cutflow_BR_mass}
and \ref{tab:cutflow_BR_lifetime}} in \suppMaterialii.  All bounds are derived at 95\% confidence
 level (CL) according to the CL$_\mathrm{s}$
prescription~\cite{CLs1,Read:2002hq,LHCCLs,Cowan:2010js} in the asymptotic approximation.
For each limit derivation, we consider events with $N_{\text{tags}}\geq 2$,
using independent bins for $N_{\text{tags}}=2$ and
$N_{\text{tags}}\geq 3$. Finer binning of the tag multiplicity for $N_{\text{tags}}>3$
is found to have a negligible effect on the expected limits.
Cross section upper limits are
 presented as a function of the
mass and proper decay length of the parent particle.  The analysis sensitivity is
maximal for $c\tau_0$ ranging from 10 to 1000\unit{mm}. Mass exclusion bounds at
fixed decay length are also derived by comparing the excluded cross section
with the values predicted for the benchmark models described in
Section~\ref{sec:samples}. In the case of SUSY models, the
next-to-leading order (NLO) and next-to-leading logarithmic (NLL)
$\PSQt\,\PSQt^*$ production cross section computed in the large-mass limit for all other SUSY particles~\cite{NLONLL1,NLONLL2,NLONLL3,NLONLL4,NLONLL5,NLONLLerr} is used as a reference.

\begin{figure*}[htbp]
\centering
\includegraphics[width=.45\textwidth]{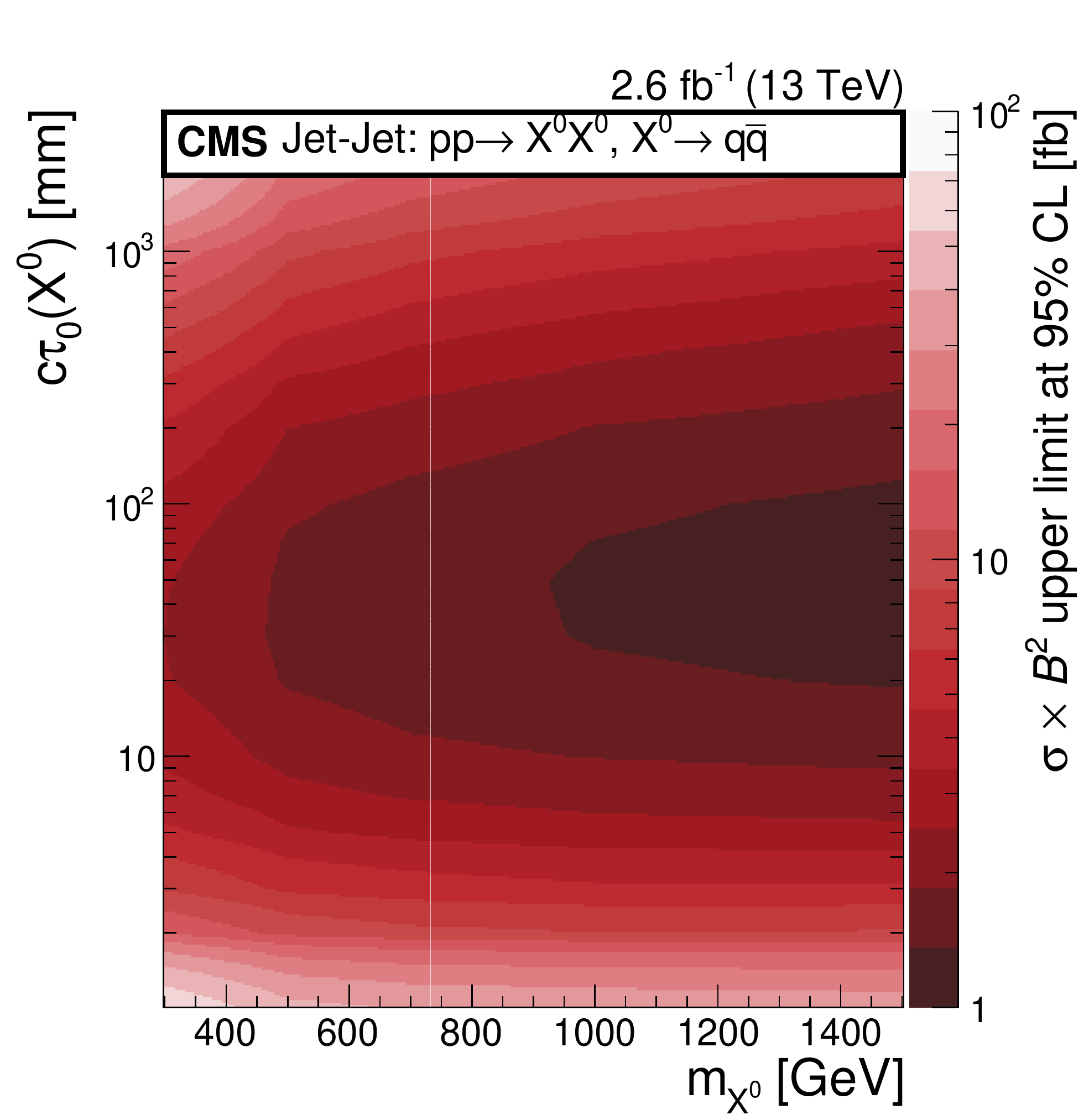}
\includegraphics[width=.45\textwidth]{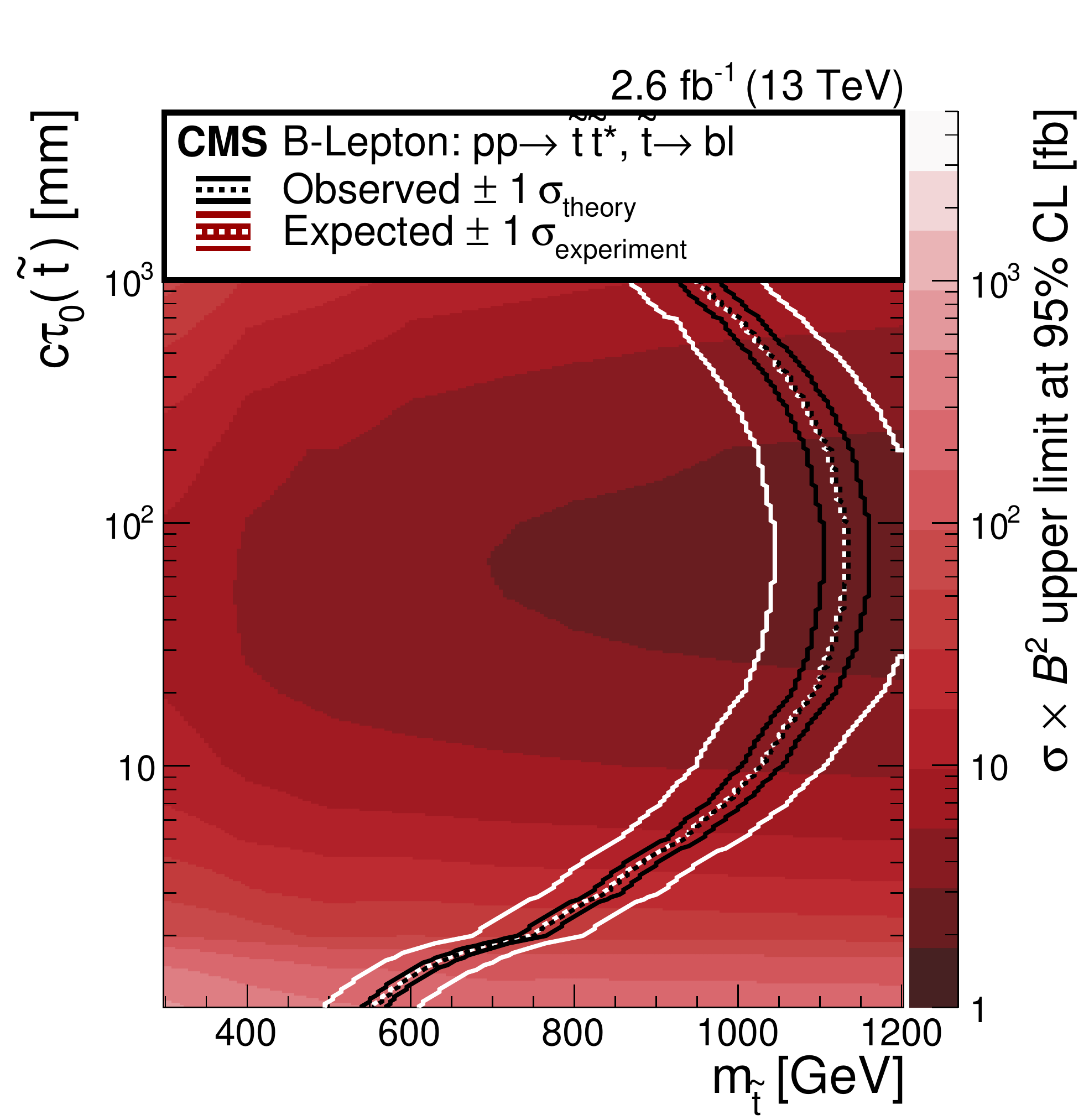}
\includegraphics[width=.7\textwidth]{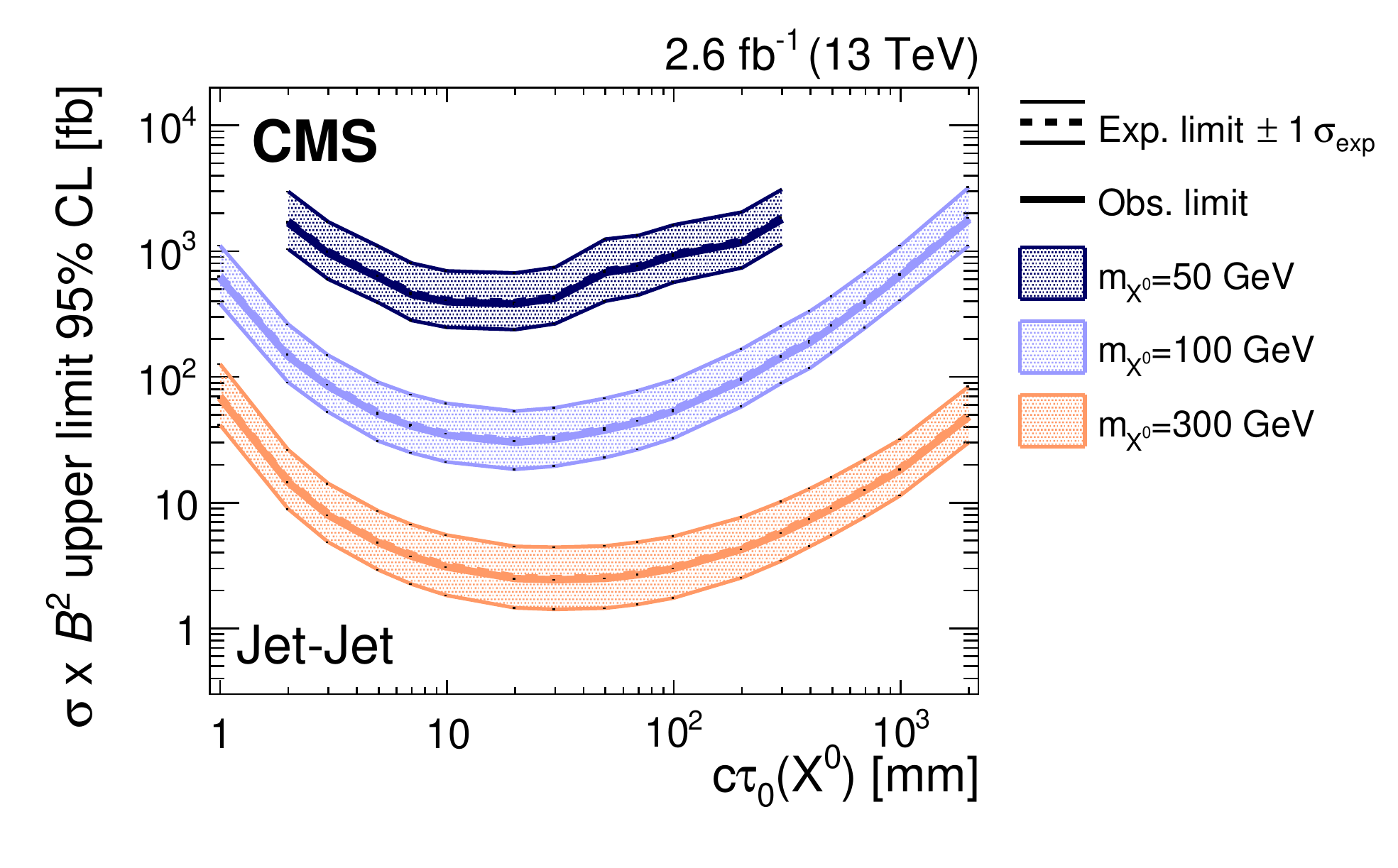}
\caption{The excluded cross section at 95\% CL for the Jet-Jet model (upper left) and the B-Lepton model
(upper right) as a function of the mass and proper decay length of the parent particle. The B-Lepton plot
 also shows the expected (observed) exclusion
 region with one standard deviation experimental (theoretical) uncertainties,
 utilizing a NLO+NLL calculation of the top squark production cross section.
The lower plot shows the excluded cross section at 95\% CL for the Jet-Jet model
as a function of the proper decay length for three illustrative smaller
values of the mass. The shaded bands in the lower plot represent the one standard
 deviation uncertainties in the expected limits.
  \label{fig:dsusy_xx4j_limit}}
\end{figure*}

Figure~\ref{fig:dsusy_xx4j_limit} shows the excluded pair production cross section for the Jet-Jet
and B-Lepton models. The Light-Light model is shown in Figure~\ifthenelse{\boolean{cms@external}}{1}{\ref{fig:light_limit}}
of \suppMaterialii and has nearly identical
performance to the Jet-Jet model. The B-Lepton sensitivity is similar to that observed for the Jet-Jet model,
although it is less stringent as additional jets give higher efficiency than
additional leptons from both the tagging and triggering
perspectives. Cross sections larger than 2.5\unit{fb} are excluded at 95\%
CL, for $c\tau_0$ in the range 70--100\unit{mm}, which corresponds to
the exclusion of parent masses below 1130\GeV.

The exclusions for the B-Tau, B-Electron and B-Muon models are shown in Figs.~\ifthenelse{\boolean{cms@external}}{2--4}{\ref{fig:dsusy_limit_tau}, \ref{fig:dsusy_limit_ele}, and \ref{fig:dsusy_limit_mu}} of \suppMaterialii, respectively.  The B-Tau and B-Electron
models have similar performance at high mass  with slightly stronger limits
for the B-Electron model at lower
mass $(m_{\sTop}=300\GeV)$ and longer decay length $(c\tau_0>10\mm)$. The
highest mass excluded in the B-Electron (B-Tau) model is
$m_{\sTop}=1145\,(1150)\GeV$ at $c\tau_0=70\mm$, corresponding to a cross section
 of 2.3\,(2.2)\unit{fb} at 95\% CL.

In the case of the B-Muon model, the analysis uses jets reconstructed from calorimetric
deposits and the two muons have small or no associated calorimeter deposits, thus the signal
reconstruction efficiency and the displaced-jet multiplicity are smaller. This results in a
weaker exclusion bound.  The highest mass excluded in the B-Muon model
is $m_{\sTop}=1085\GeV$ at $c\tau_0=70\mm$, corresponding to a cross
section upper limit of 3.5\unit{fb} at 95\% CL.

\section{Summary}
\label{sec:conclusion}

A search for long-lived particles has been performed with data corresponding to
 an integrated luminosity of 2.6\fbinv collected at a center-of-mass energy of
13\TeV by the CMS experiment in 2015. This is the first search for
long-lived particles decaying to jet final states in 13\TeV data and the first
search to demonstrate explicit sensitivity to long-lived particles decaying to $\tau$ leptons.
  The analysis utilizes two customized topological
 trigger algorithms and an offline displaced-jet tagging algorithm, where
the multiplicity of displaced jets is used to search for the presence of a signal.
  As no excess above the predicted background is found, upper limits are set
 at 95\% confidence level on the production cross section for long-lived
resonances decaying to two jets or to a lepton and b quark.  The limits are calculated as a
 function of the mass and proper decay length of the long-lived particles.
For resonances decaying to a b quark and a lepton, cross sections larger than
2.5\unit{fb} are excluded for proper decay lengths of 70--100 \mm.
 The cross section limits are also translated into mass exclusion bounds,
using a calculation of the top squark production cross section as a reference.
Assuming equal lepton branching fractions, pair-produced long-lived R-parity violating
top squarks lighter than 550--1130\GeV are excluded, depending on the squark proper decay length.
 This mass exclusion bound is currently the most stringent bound available
for top squark proper decay lengths greater than 3\mm.

\ifthenelse{\boolean{cms@external}}{}{\enlargethispage{3ex}}

\begin{acknowledgments}
We congratulate our colleagues in the CERN accelerator departments for the excellent performance of the LHC and thank the technical and administrative staffs at CERN and at other CMS institutes for their contributions to the success of the CMS effort. In addition, we gratefully acknowledge the computing center and personnel of the Worldwide LHC Computing Grid for delivering so effectively the computing infrastructure essential to our analyses. Finally, we acknowledge the enduring support for the construction and operation of the LHC and the CMS detector provided by the following funding agencies: BMWFW and FWF (Austria); FNRS and FWO (Belgium); CNPq, CAPES, FAPERJ, and FAPESP (Brazil); MES (Bulgaria); CERN; CAS, MoST, and NSFC (China); COLCIENCIAS (Colombia); MSES and CSF (Croatia); RPF (Cyprus); SENESCYT (Ecuador); MoER, ERC IUT, and ERDF (Estonia); Academy of Finland, MEC, and HIP (Finland); CEA and CNRS/IN2P3 (France); BMBF, DFG, and HGF (Germany); GSRT (Greece); OTKA and NIH (Hungary); DAE and DST (India); IPM (Iran); SFI (Ireland); INFN (Italy); MSIP and NRF (Republic of Korea); LAS (Lithuania); MOE and UM (Malaysia); BUAP, CINVESTAV, CONACYT, LNS, SEP, and UASLP-FAI (Mexico); MBIE (New Zealand); PAEC (Pakistan); MSHE and NSC (Poland); FCT (Portugal); JINR (Dubna); MON, RosAtom, RAS, RFBR and RAEP (Russia); MESTD (Serbia); SEIDI, CPAN, PCTI and FEDER (Spain); Swiss Funding Agencies (Switzerland); MST (Taipei); ThEPCenter, IPST, STAR, and NSTDA (Thailand); TUBITAK and TAEK (Turkey); NASU and SFFR (Ukraine); STFC (United Kingdom); DOE and NSF (USA).

\hyphenation{Rachada-pisek} Individuals have received support from the Marie-Curie program and the European Research Council and EPLANET (European Union); the Leventis Foundation; the A. P. Sloan Foundation; the Alexander von Humboldt Foundation; the Belgian Federal Science Policy Office; the Fonds pour la Formation \`a la Recherche dans l'Industrie et dans l'Agriculture (FRIA-Belgium); the Agentschap voor Innovatie door Wetenschap en Technologie (IWT-Belgium); the Ministry of Education, Youth and Sports (MEYS) of the Czech Republic; the Council of Science and Industrial Research, India; the HOMING PLUS programme of the Foundation for Polish Science, cofinanced from European Union, Regional Development Fund, the Mobility Plus programme of the Ministry of Science and Higher Education, the National Science Center (Poland), contracts Harmonia 2014/14/M/ST2/00428, Opus 2014/13/B/ST2/02543, 2014/15/B/ST2/03998, and 2015/19/B/ST2/02861, Sonata-bis 2012/07/E/ST2/01406; the National Priorities Research Program by Qatar National Research Fund; the Programa Clar\'in-COFUND del Principado de Asturias; the Thalis and Aristeia programmes cofinanced by EU-ESF and the Greek NSRF; the Rachadapisek Sompot Fund for Postdoctoral Fellowship, Chulalongkorn University and the Chulalongkorn Academic into Its 2nd Century Project Advancement Project (Thailand); and the Welch Foundation, contract C-1845.

\end{acknowledgments}

\bibliography{auto_generated}
\ifthenelse{\boolean{cms@external}}{}{
\numberwithin{table}{section}
\numberwithin{figure}{section}
\clearpage
\appendix
\section{Appendix \label{app:suppMat}}

\begin{table*}[hbt]
  \topcaption{ Signal efficiencies (in \%) for  $c\tau_0=30$\mm and
   various values of mass with modified branching ratios relative
   to the Jet-Jet and B-Lepton models. Selection requirements are cumulative from
    the first row to the last.\label{tab:cutflow_BR_mass}}
\centering
\begin{tabular}{lccccc}
\multicolumn{6}{c}{{Light-Light}} \\
 \hline
 $m_{X^0}\,[\GeVns]$ & 50 & 100 & 300 & 1000 & 1500 \\
 \hline
 $\geq$2  tags & 2.84 $\pm$ 0.12   & 15.56 $\pm$ 0.29  & 54.87 $\pm$ 0.92 & 80.52 $\pm$ 1.11 & 82.19 $\pm$ 1.14 \\
 Trigger       & 0.53 $\pm$ 0.05 & 5.70 $\pm$ 0.17   & 47.14 $\pm$ 0.85 & 74.85 $\pm$ 1.07 & 77.07 $\pm$ 1.10 \\
 Selection     & 0.33 $\pm$ 0.04 & 3.90 $\pm$ 0.14   & 45.68 $\pm$ 0.84 & 74.80 $\pm$ 1.07 & 76.96 $\pm$ 1.10 \\
 $\geq$3  tags & 0.05 $\pm$ 0.02 & 1.11 $\pm$ 0.08   & 21.77 $\pm$ 0.58 & 50.04 $\pm$ 0.88 & 55.36 $\pm$ 0.93 \\
 $\geq$4  tags & \NA                & 0.23 $\pm$ 0.04 & 7.38 $\pm$ 0.34  & 25.80 $\pm$ 0.63 & 32.47 $\pm$ 0.71 \\
\end{tabular}

\vspace*{3ex}
\begin{tabular}{lcccc}
\multicolumn{5}{c}{{B-Electron}} \\
 \hline
 $m_{\sTop}\,[\GeVns]$ & 300 & 600 & 800 & 1000 \\
\hline
 $\geq$2  tags     & 39.01 $\pm$ 0.65 & 53.70 $\pm$ 0.75 & 59.62 $\pm$ 0.78 & 62.42 $\pm$ 1.11 \\
         Trigger          & 22.95 $\pm$ 0.50 & 38.07 $\pm$ 0.63 & 43.06 $\pm$ 0.66 & 45.21 $\pm$ 0.95 \\
          Selection       & 21.59 $\pm$ 0.48 & 37.02 $\pm$ 0.62 & 39.47 $\pm$ 0.64 & 42.20 $\pm$ 0.92 \\
 $\geq$3  tags  & 7.86 $\pm$ 0.29  & 14.28 $\pm$ 0.38 & 17.37 $\pm$ 0.42 & 20.39 $\pm$ 0.64 \\
 $\geq$4  tags & 1.37 $\pm$ 0.12  & 3.32 $\pm$ 0.19  & 4.34 $\pm$ 0.21  & 4.69 $\pm$ 0.31  \\
\end{tabular}
\vspace*{3ex}
\begin{tabular}{lcccc}
\multicolumn{5}{c}{{B-Tau}} \\
 \hline
 $m_{\sTop}\,[\GeVns]$ & 300 & 600 & 800 & 1000 \\
 \hline
$\geq 2$ tags  & 34.98 $\pm$ 0.61 & 51.42 $\pm$ 0.73 & 57.20 $\pm$ 0.76 & 59.43 $\pm$ 1.07 \\
 Trigger       & 20.20 $\pm$ 0.46 & 39.78 $\pm$ 0.64 & 45.46 $\pm$ 0.68 & 47.62 $\pm$ 0.96 \\
 Selection     & 17.17 $\pm$ 0.43 & 37.47 $\pm$ 0.62 & 43.64 $\pm$ 0.67 & 44.26 $\pm$ 0.92 \\
 $\geq$3  tags & 5.21 $\pm$ 0.24  & 13.29 $\pm$ 0.37 & 16.15 $\pm$ 0.40 & 19.13 $\pm$ 0.61 \\
 $\geq$4  tags & 0.86 $\pm$ 0.10  & 3.09 $\pm$ 0.18  & 3.68 $\pm$ 0.19  & 4.48 $\pm$ 0.29  \\
\end{tabular}
\vspace*{3ex}
\begin{tabular}{lcccc}
\multicolumn{5}{c}{{B-Muon}} \\
 \hline
 $m_{\sTop}\,[\GeVns]$ & 300 & 600 & 800 & 1000 \\
\hline
$\geq 2$ tags & 20.09 $\pm$ 0.46  & 35.46 $\pm$ 0.60  & 41.18 $\pm$ 0.64  & 43.13 $\pm$ 0.93 \\
Trigger       & 6.63 $\pm$ 0.26   & 24.73 $\pm$ 0.50  & 31.85 $\pm$ 0.56  & 34.10 $\pm$ 0.82 \\
Selection     & 5.25 $\pm$ 0.24   & 21.40 $\pm$ 0.47  & 27.42 $\pm$ 0.52  & 31.18 $\pm$ 0.79 \\
$\geq 3$ tags & 0.34 $\pm$ 0.06 & 3.03 $\pm$ 0.18   & 5.28 $\pm$ 0.23   & 6.08 $\pm$ 0.35  \\
$\geq 4$ tags & \NA                & 0.12 $\pm$ 0.04 & 0.68 $\pm$ 0.08 & 0.68 $\pm$ 0.12  \\
\end{tabular}
\end{table*}

\begin{table*}[tbhp!]
  \caption{ Signal efficiencies (in \%) for $m_{X^0}=m_{\sTop}=300$\GeV
    and for various values of $c\tau_0$ with modified branching ratios relative to
    the Jet-Jet and B-Lepton models. Selection requirements are cumulative from
    the first row to the last.
    \label{tab:cutflow_BR_lifetime}}
\centering
\begin{tabular}{lcccc}
\multicolumn{5}{c}{{Light-Light}} \\
 \hline
 $c\tau_0 \,[\text{mm}]$ & 1 & 10 & 100 & 1000 \\
 \hline
 $\geq$2  tags & 2.20 $\pm$ 0.19   & 40.49 $\pm$ 0.80 & 54.92 $\pm$ 0.93 & 14.55 $\pm$ 0.47  \\
 Trigger       & 2.04 $\pm$ 0.18   & 39.16 $\pm$ 0.78 & 39.63 $\pm$ 0.79 & 8.20 $\pm$ 0.36   \\
 Selection     & 2.03 $\pm$ 0.18   & 38.41 $\pm$ 0.77 & 36.99 $\pm$ 0.76 & 6.89 $\pm$ 0.33   \\
 $\geq$3  tags & 0.19 $\pm$ 0.05 & 14.77 $\pm$ 0.48 & 16.70 $\pm$ 0.51 & 1.48 $\pm$ 0.15   \\
 $\geq$4  tags & \NA                & 5.11 $\pm$ 0.28  & 4.73 $\pm$ 0.27  & 0.22 $\pm$ 0.06 \\
\end{tabular}

\vspace*{3ex}
\begin{tabular}{lcccc}
\multicolumn{5}{c}{{B-Electron}} \\
 \hline
 $c\tau_0 \,[\text{mm}]$ & 1 & 10 & 100 & 1000 \\
 \hline
 $\geq$2  tags     & 0.81 $\pm$ 0.10 & 20.51 $\pm$ 0.47  & 39.01 $\pm$ 0.65 & 11.46 $\pm$ 0.35  \\
   Trigger          & 0.40 $\pm$ 0.07 & 14.68 $\pm$ 0.40  & 22.95 $\pm$ 0.50 & 5.15 $\pm$ 0.23   \\
    Selection       & 0.40 $\pm$ 0.07 & 13.92 $\pm$ 0.39  & 20.34 $\pm$ 0.47 & 3.58 $\pm$ 0.19   \\
 $\geq$3  tags  & 0.04 $\pm$ 0.02 & 4.22 $\pm$ 0.21   & 7.21 $\pm$ 0.28  & 0.82 $\pm$ 0.09 \\
 $\geq$4  tags & \NA                & 0.73 $\pm$ 0.09 & 1.19 $\pm$ 0.11  & 0.05 $\pm$ 0.02 \\
\end{tabular}
\vspace*{3ex}
\begin{tabular}{lcccc}
\multicolumn{5}{c}{{B-Tau}} \\
 \hline
 $c\tau_0\,[\text{mm}]$ & 1 & 10 & 100 & 1000 \\
\hline
$\geq$2 tags & 0.48 $\pm$ 0.07 & 18.40 $\pm$ 0.45  & 34.98 $\pm$ 0.61  & 9.31 $\pm$ 0.32   \\
Trigger       & 0.44 $\pm$ 0.07 & 14.63 $\pm$ 0.40  & 20.20 $\pm$ 0.46  & 3.81 $\pm$ 0.20   \\
Selection     & 0.41 $\pm$ 0.07 & 12.45 $\pm$ 0.37  & 15.50 $\pm$ 0.41  & 2.37 $\pm$ 0.16   \\
$\geq$3 tags & 0.02 $\pm$ 0.02 & 3.23 $\pm$ 0.19   & 4.62 $\pm$ 0.22   & 0.44 $\pm$ 0.07 \\
$\geq$4 tags & \NA                & 0.53 $\pm$ 0.08 & 0.66 $\pm$ 0.09   & 0.02 $\pm$ 0.02 \\
\end{tabular}
\vspace*{3ex}
\begin{tabular}{lcccc}
\multicolumn{5}{c}{{B-Muon}} \\
 \hline
 $c\tau_0\,[\text{mm}]$ & 1 & 10 & 100 & 1000 \\
 \hline
 $\geq$2  tags     & 0.13 $\pm$ 0.04 & 8.02 $\pm$ 0.29   & 20.09 $\pm$ 0.46  & 4.03 $\pm$ 0.21   \\
   Trigger          & 0.05 $\pm$ 0.02 & 3.97 $\pm$ 0.21   & 6.63 $\pm$ 0.26   & 0.88 $\pm$ 0.10 \\
    Selection       & 0.04 $\pm$ 0.02 & 2.92 $\pm$ 0.18   & 4.21 $\pm$ 0.21   & 0.49 $\pm$ 0.07 \\
 $\geq$3  tags  & \NA                & 0.23 $\pm$ 0.05 & 0.31 $\pm$ 0.06 & 0.03 $\pm$ 0.02 \\
 $\geq$4  tags & \NA                & 0.01 $\pm$ 0.01 & \NA                & \NA                \\
\end{tabular}
\end{table*}

\begin{figure*}
\centering
\resizebox{\textwidth}{!}{{\includegraphics[height=20mm]{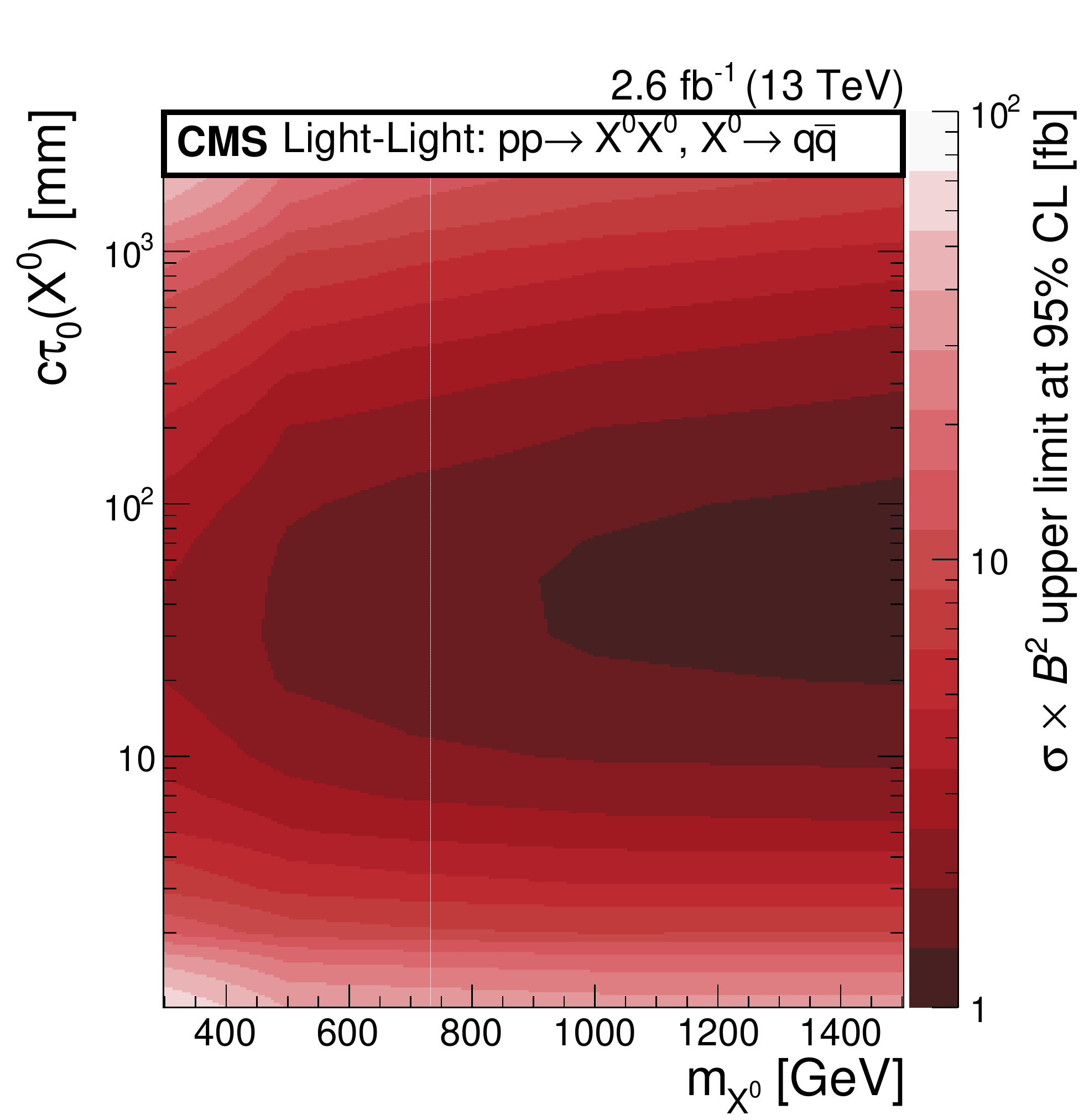}}
{\includegraphics[height=20mm]{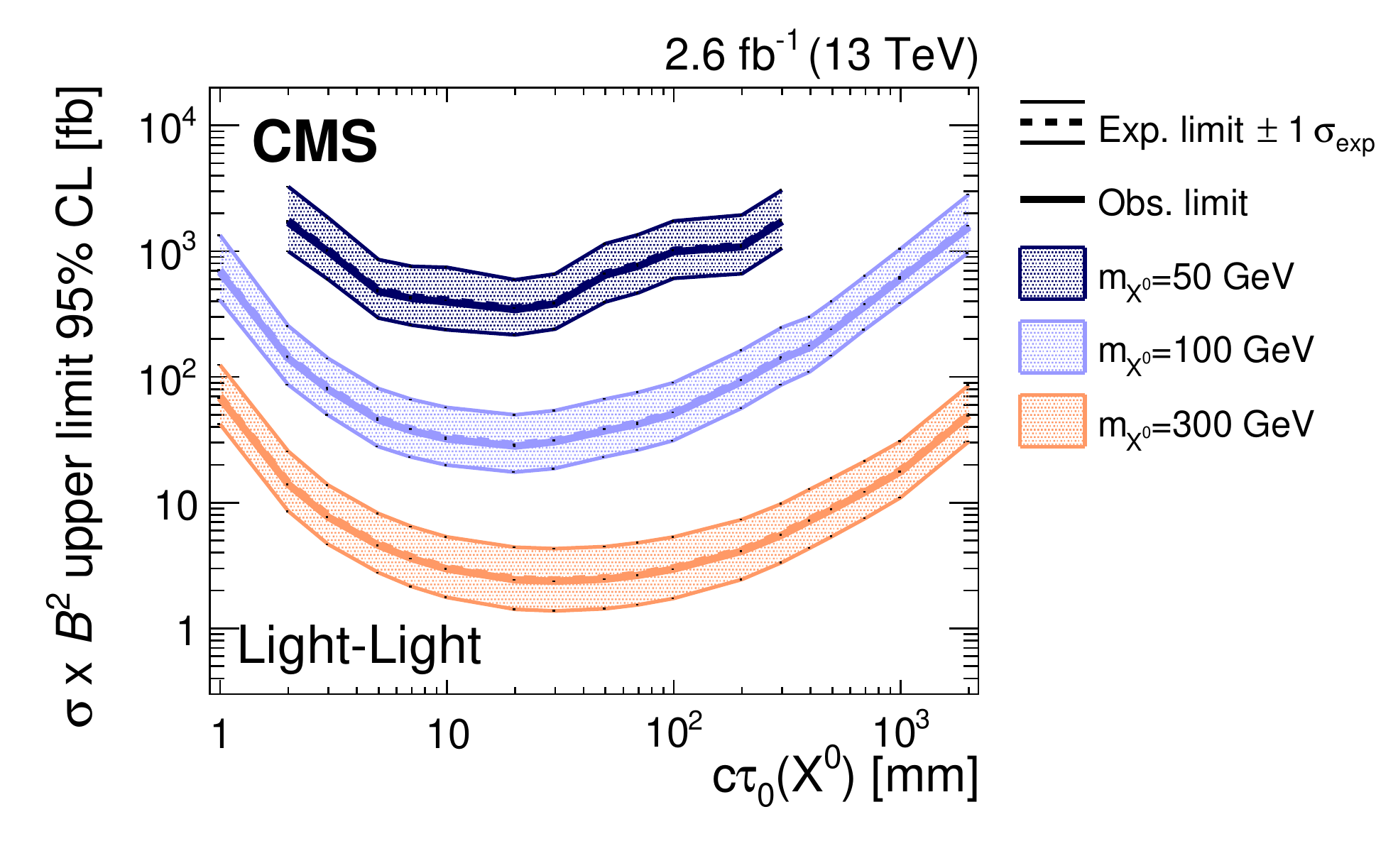}}}
\caption{The excluded cross section at 95\% CL for the Light-Light
  model as a function of the mass and proper decay length of the parent particle
  $X^0$ (left) and as a function of the proper decay length for three illustrative smaller values of the
  mass (right). The shaded bands in the right plot represent the one standard
  deviation uncertainties in the expected limits. \label{fig:light_limit}}
\end{figure*}

\begin{figure*}
\centering
\resizebox{\textwidth}{!}{{\includegraphics[height=20mm]{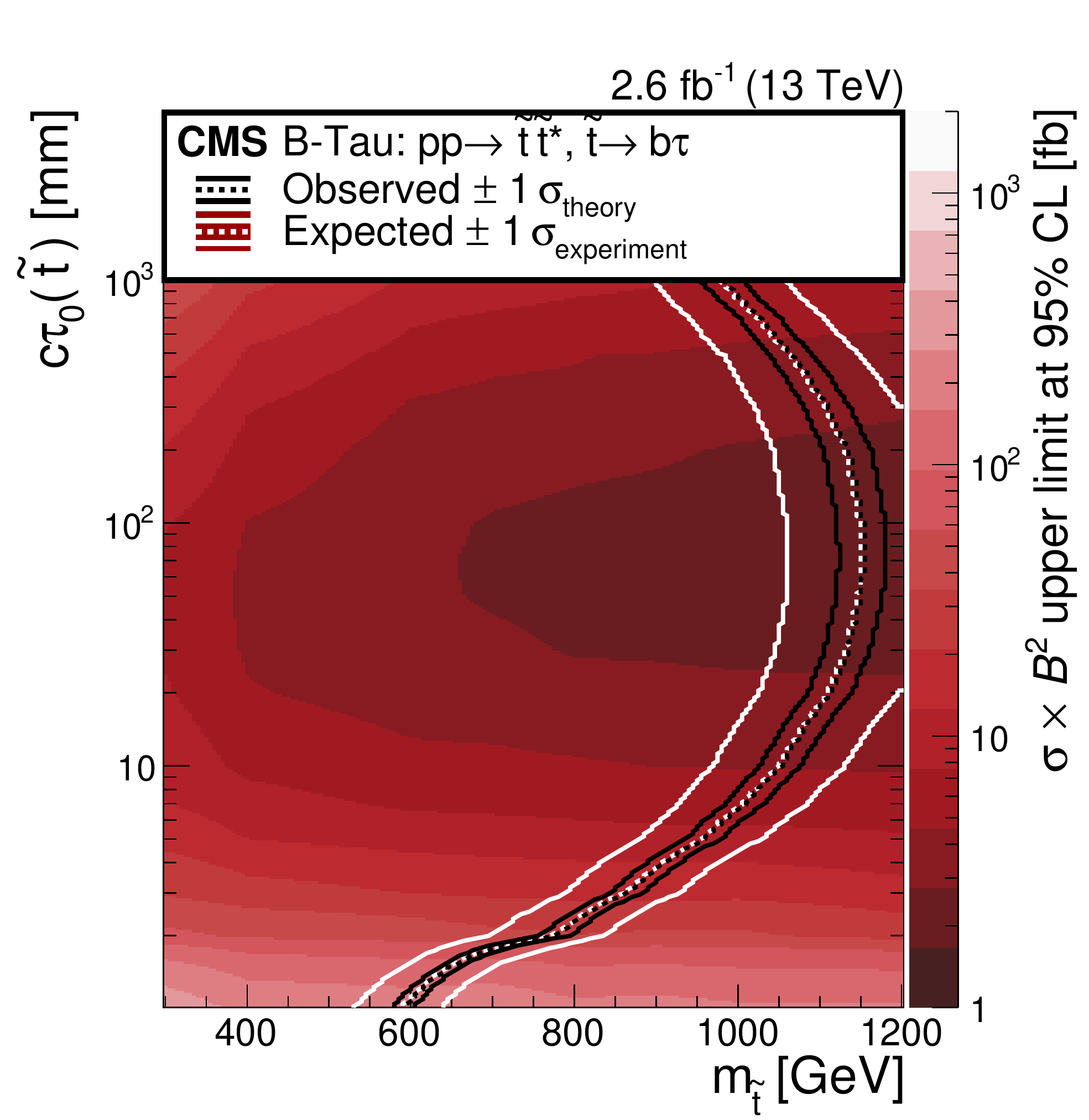}}
{\includegraphics[height=20mm]{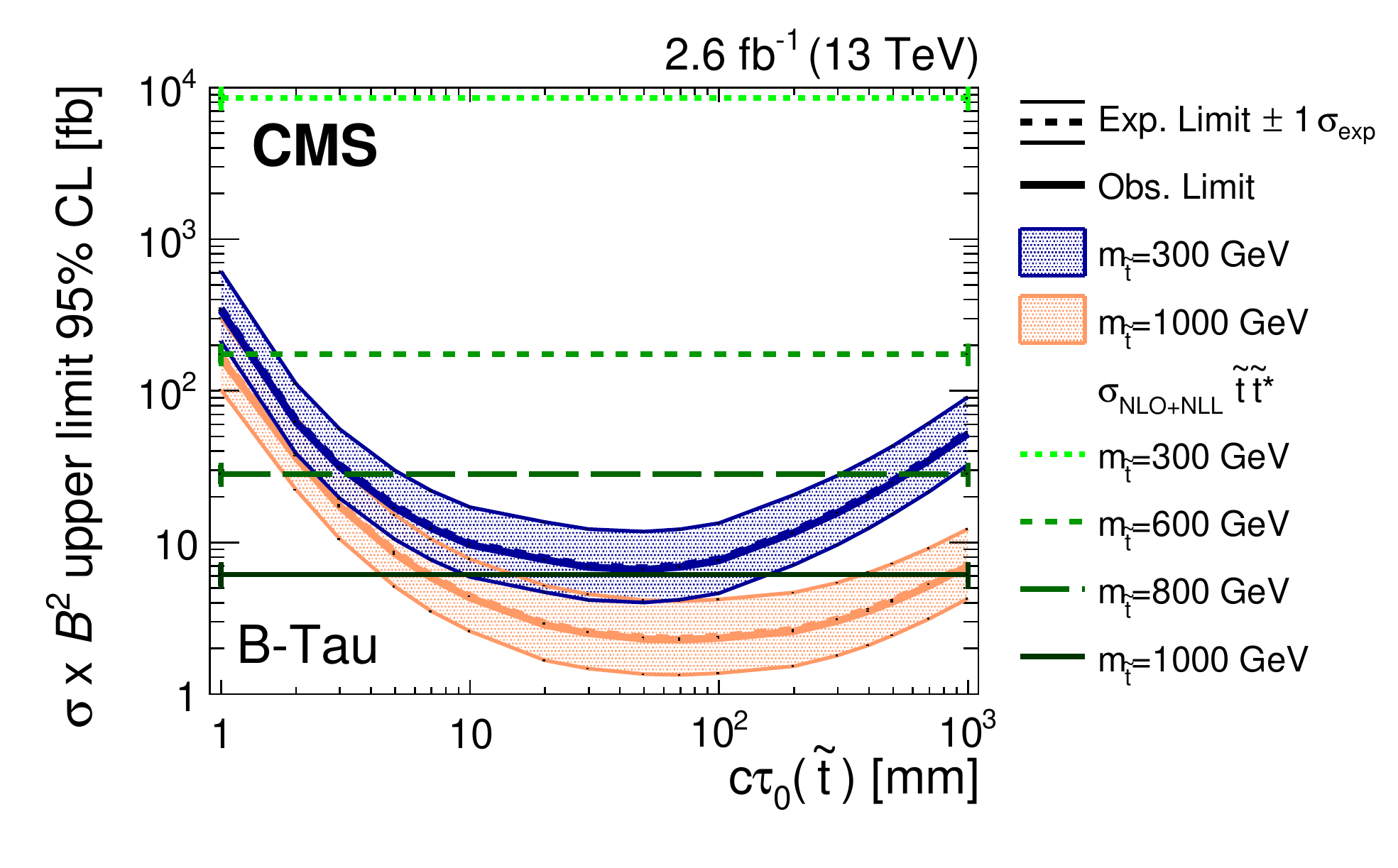}}}\caption{The excluded cross section at 95\% CL for the B-Tau model as
  a function of the mass and proper decay length of the parent particle
  $\sTop$ (left) and as a function of the proper decay length for two values
  of the mass (right).  The left plot also shows the expected (observed) exclusion
 region with one standard deviation experimental (theoretical) uncertainties,
 utilizing a NLO+NLL calculation of the top squark production cross section.
  The right plot also shows the expected left limits with one standard
 deviation uncertainties as bands. The NLO+NLL calculation of the top squark
 production cross section is drawn horizontally in green for four mass values.
\label{fig:dsusy_limit_tau}}
\end{figure*}

\begin{figure*}
\centering
\resizebox{\textwidth}{!}{{\includegraphics[height=20mm]{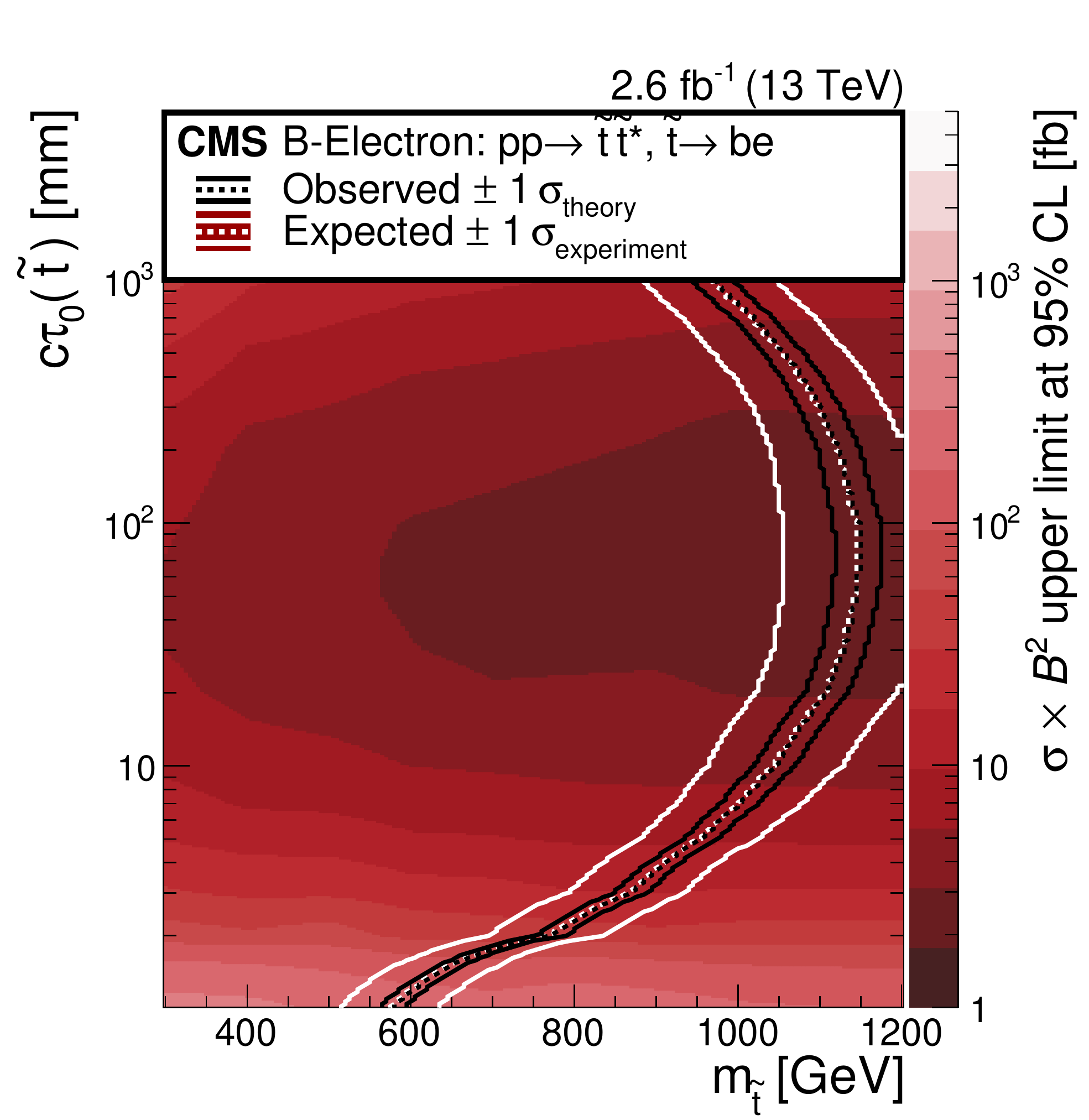}}
{\includegraphics[height=20mm]{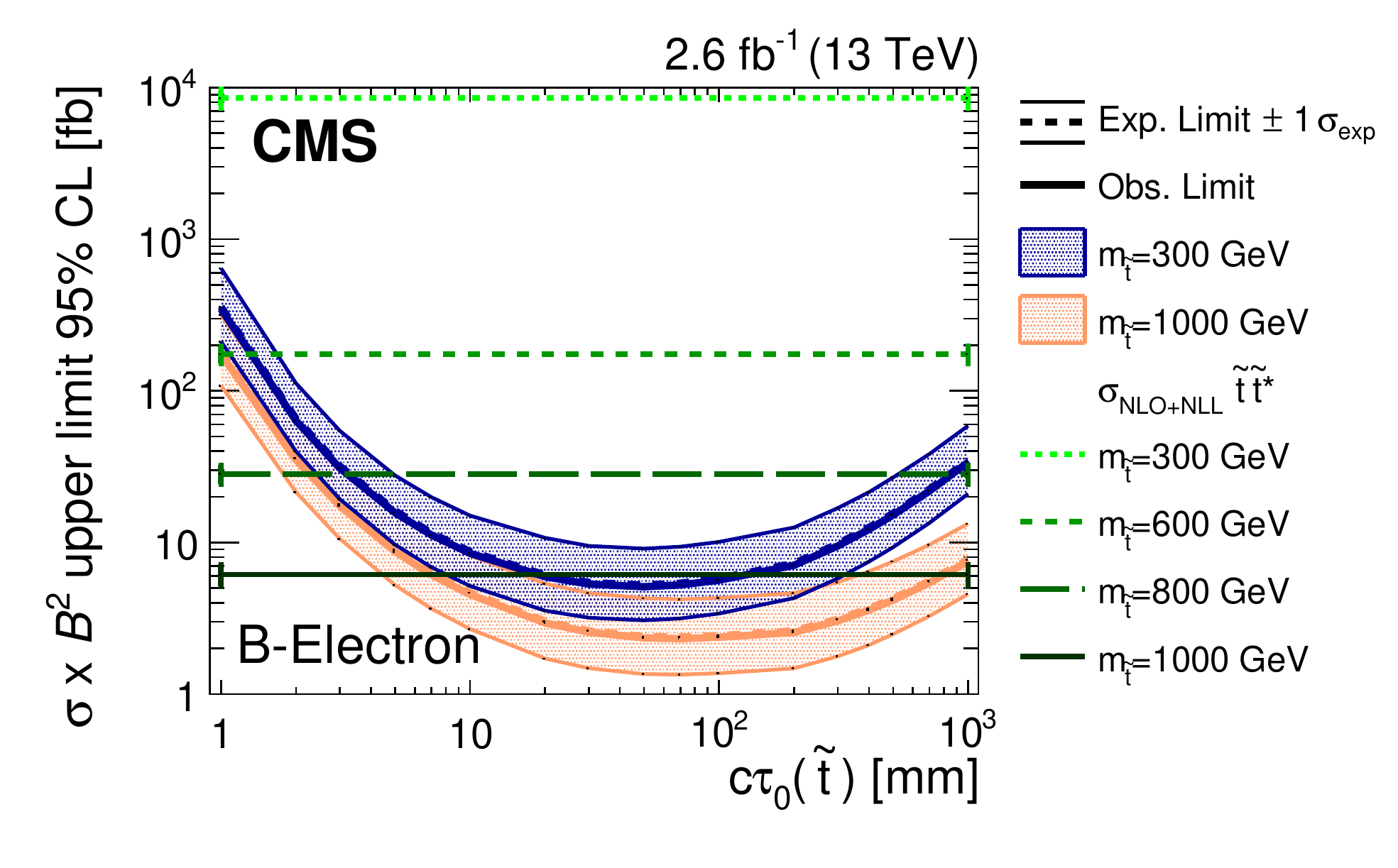}}}\caption{ The excluded cross section at 95\% CL for the B-Electron model as
  a function of the mass and proper decay length of the parent particle
  $\sTop$ (left) and as a function of the proper decay length for two values
  of the mass (right).  The left plot also shows the expected (observed) exclusion
 region with one standard deviation experimental (theoretical) uncertainties,
 utilizing a NLO+NLL calculation of the top squark production cross section.
  The right plot also shows the expected left limits with one standard
 deviation uncertainties as bands. The NLO+NLL calculation of the top squark
 production cross section is drawn horizontally in green for four mass values.
\label{fig:dsusy_limit_ele}}
\end{figure*}

\begin{figure*}
\centering
\resizebox{\textwidth}{!}{{\includegraphics[height=20mm]{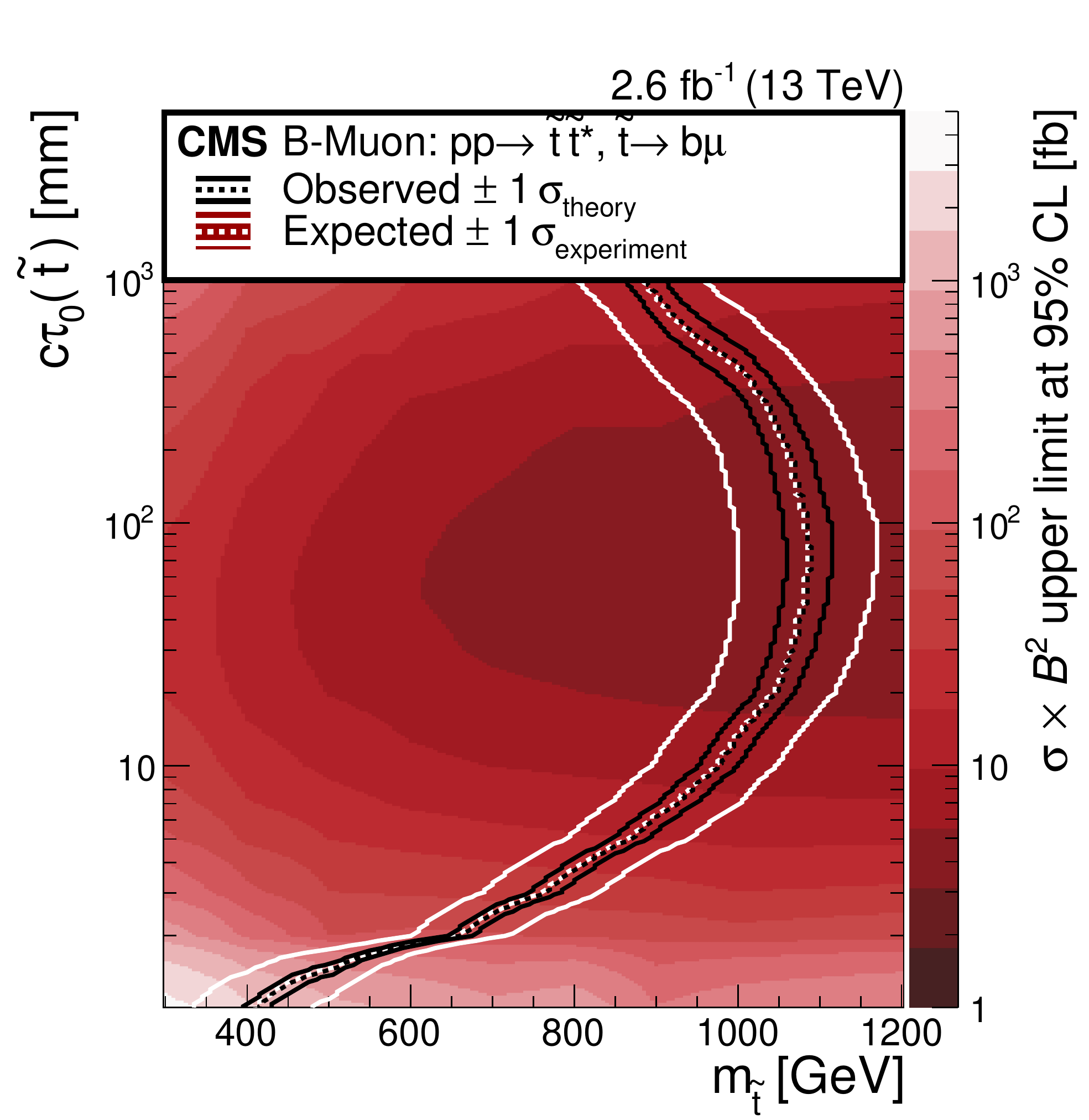}}
{\includegraphics[height=20mm]{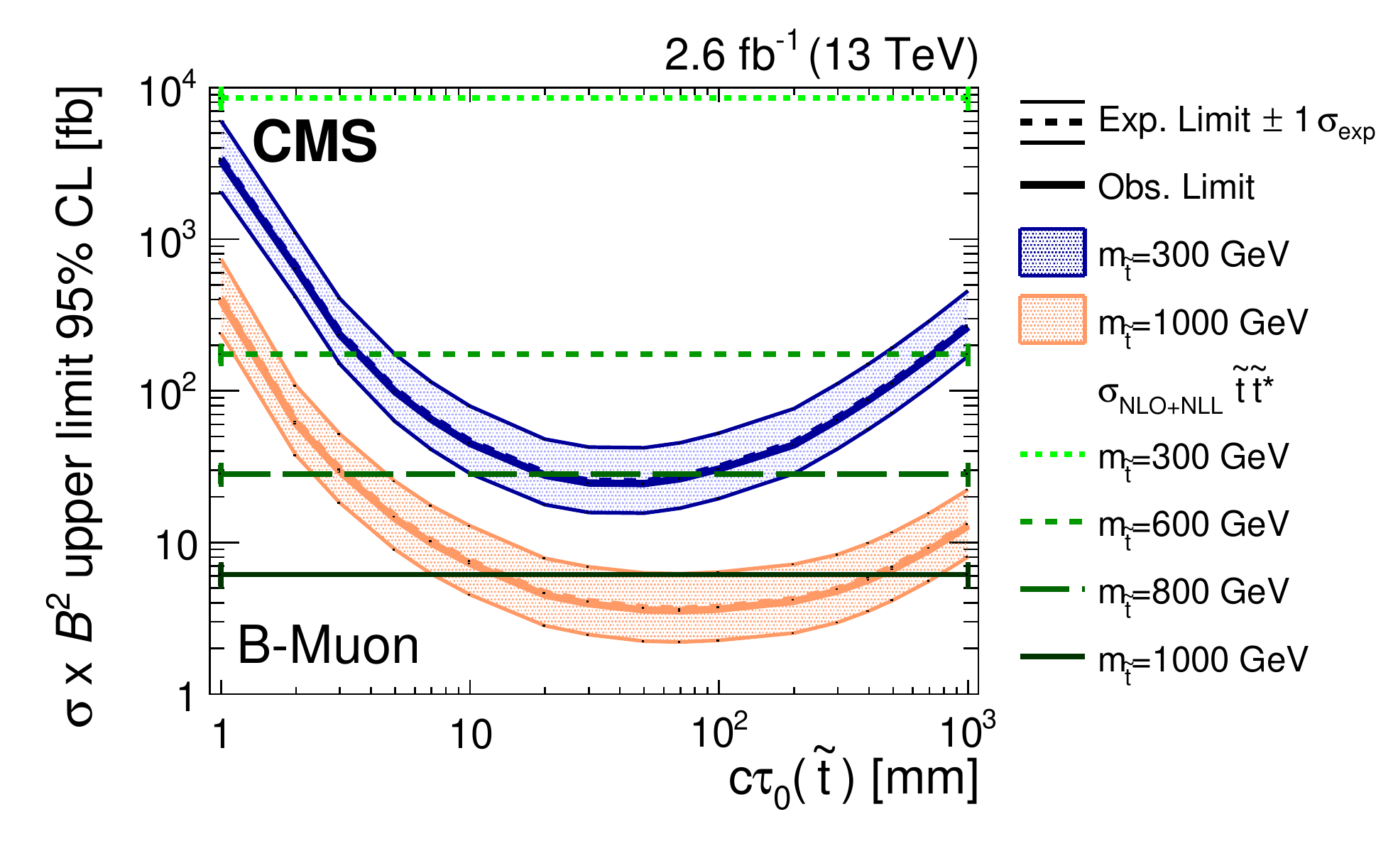}}}\caption{ The excluded cross section at 95\% CL for the B-Muon model as
  a function of the mass and proper decay length of the parent particle
  $\sTop$ (left) and as a function of the proper decay length for two values
  of the mass (right).  The left plot also shows the expected (observed) exclusion
 region with one standard deviation experimental (theoretical) uncertainties,
 utilizing a NLO+NLL calculation of the top squark production cross section.
  The right plot also shows the expected left limits with one standard
 deviation uncertainties as bands. The NLO+NLL calculation of the top squark
 production cross section is drawn horizontally in green for four mass values.
\label{fig:dsusy_limit_mu}}
\end{figure*}

}
\cleardoublepage \section{The CMS Collaboration \label{app:collab}}\begin{sloppypar}\hyphenpenalty=5000\widowpenalty=500\clubpenalty=5000\textbf{Yerevan Physics Institute,  Yerevan,  Armenia}\\*[0pt]
A.M.~Sirunyan, A.~Tumasyan
\vskip\cmsinstskip
\textbf{Institut f\"{u}r Hochenergiephysik,  Wien,  Austria}\\*[0pt]
W.~Adam, F.~Ambrogi, E.~Asilar, T.~Bergauer, J.~Brandstetter, E.~Brondolin, M.~Dragicevic, J.~Er\"{o}, M.~Flechl, M.~Friedl, R.~Fr\"{u}hwirth\cmsAuthorMark{1}, V.M.~Ghete, J.~Grossmann, J.~Hrubec, M.~Jeitler\cmsAuthorMark{1}, A.~K\"{o}nig, N.~Krammer, I.~Kr\"{a}tschmer, D.~Liko, T.~Madlener, I.~Mikulec, E.~Pree, D.~Rabady, N.~Rad, H.~Rohringer, J.~Schieck\cmsAuthorMark{1}, R.~Sch\"{o}fbeck, M.~Spanring, D.~Spitzbart, J.~Strauss, W.~Waltenberger, J.~Wittmann, C.-E.~Wulz\cmsAuthorMark{1}, M.~Zarucki
% ...
\vskip\cmsinstskip
% ...
1:~~Also at Vienna University of Technology, Vienna, Austria\\
% ...

\end{sloppypar}
\end{document}